\documentclass[
aps,
nofootinbib,
superscriptaddress,
tightenlines,
notitlepage,
twocolumn,
showpacs,
floatfix,
]{revtex4-1}
\usepackage{amssymb,amsmath,bm,tensor,braket}
\usepackage[colorlinks]{hyperref}

\usepackage{graphicx}
\usepackage{epstopdf}
\graphicspath{{figures/}}
\usepackage{dcolumn}
\usepackage{bm}
\usepackage{xcolor}

\newcommand{\BIT}{\affiliation{School of Physics, Beijing Institute of Technology, Beijing, 100081, China}}

\begin{document}

\preprint{APS/123-QED}

\title{Accreting Black Holes in Dark Matter Halos}

\author{Sobhan Kazempour}\email{sobhan.kazempour1989@gmail.com}\BIT
\author{Sichun Sun}\email{sichunssun@gmail.com}\BIT
\author{Chengye Yu}\email{chengyeyu1@hotmail.com}\BIT

\begin{abstract}
We examine the thin accretion disk behaviors surrounding black holes embedded in cold dark matter halos and scalar field dark matter halos. We first calculate the event horizons and derive the equations of motion and effective potential in black hole geometries with different dark matter halos. We then compute the specific energy, specific angular momentum, and angular velocity of particles moving along circular orbits. We also derive the effective potentials to find the locations of the innermost stable circular orbit (ISCO) and compare them to the Schwarzschild and Kerr black holes without the dark matter halos. We also use the observed ISCO of the supermassive black hole at the Galactic Center of the Milky Way, Sagittarius A*, to constrain the dark matter halos.
\end{abstract}


\maketitle


\section{Introduction}\label{sec:1}

Einstein's theory of general relativity has made numerous accurate predictions from black holes to gravitational waves \cite{Will:2014kxa,Reynaud:2008yd,Everitt:2011hp,Berti:2015itd}. In the meantime, dark matter and dark energy need to be included to account for certain fundamental phenomena \cite{ParticleDataGroup:2014cgo,Planck:2018vyg,Peebles:2002gy}. It is important to note that the sign of dark matter is only observed through its gravitational effects so far\cite{Bertone:2004pz}. Black holes are among the successful predictions of general relativity and serve as excellent objects for studying the interplay between matter, space-time, and their gravitational interactions in astrophysical contexts. Furthermore, the accretion disks surrounding supermassive compact objects offer valuable insights into strong gravity and the underlying nature of physical processes \cite{Abramowicz:2011xu}.

In this study, we aim to provide insights into the properties of the accretion disks of rotating black holes surrounded by dark matter halos. By examining the accretion disks associated with these black holes, we hope to gain a deeper understanding of their behavior and characteristics, particularly in comparison to the well-known Schwarzschild and Kerr black hole geometries, which do not take dark matter halos into account. 

As the formation of supermassive black holes in galaxy centers remains a mystery in high-energy astrophysics, dark matter halos have been suggested as a solution, especially in the early universe \cite{Balberg:2001qg,Balberg:2002ue}. However, the nature of dark matter is still unclear, making their impact on black hole formation uncertain. Several studies have focused on the space-time geometry of dark matter halos without black holes, assuming a flat geometry due to low dark matter density and the absence of relativistic motion. This approach allowed for the study of dynamic processes like tidal disruption and stellar motion using geometric methods \cite{Matos:2003nb,Matos:2004je,Fay:2004vw}.
The strong gravity of a supermassive black hole can amplify dark matter densities, leading to the 'spike' phenomenon \cite{Fields:2014pia,Sadeghian:2013laa,Gondolo:1999ef}. 
On the other hand, the Navarro-Frenk-White density profile encounters a challenge known as the 'cusp' problem, which contrasts with observational evidence suggesting a flatter density distribution \cite{deBlok:2009sp}. Meanwhile, dark matter models like scalar field dark matter, modified Newtonian dynamics dark matter, and warm dark matter do not produce a 'cusp' on small scales. It is worth pointing out that the recent review \cite{Salucci:2018hqu}, offers a comprehensive analysis of dark matter distribution within galaxies. This review illuminates the intricate dynamics of dark matter and its impact on galactic structures, emphasizing the ongoing endeavors to unravel the mysteries surrounding dark matter and its role in shaping the universe.

It should be mentioned that recent findings have enabled the study of black holes surrounded by dark matter halos, allowing for the exploration of diverse dynamical processes and the energy density of dark matter in relativistic regimes \cite{Jusufi:2019nrn,Choi:2014hda,Carr:2020xqk,DeLuca:2023laa}. In Ref. \cite{Xu:2018wow}, the authors derived the analytical form of black hole space-time in dark matter halos, enabling the study of black hole-dark matter interactions. They provided valuable insights into the role of dark matter in shaping the black hole environment.
Furthermore, these kinds of studies open up opportunities for further exploration in astrophysics, including the examination of accretion disks surrounding black holes, the black hole shadow in dark matter halos, and the dynamic processes influenced by dark matter \cite{Pugliese:2022oes,Boshkayev:2020kle,Konoplya:2019sns,Booth:2009zb}.

Accretion disks surrounding astrophysical black holes are crucial tools for understanding strong-field gravity and testing modified theories. Black holes grow in mass through accretion, and the accompanying disk forms as diffuse material spirals toward the central object while radiating away gravitational energy. Observed spectral features of these disks, introduced by Shakura and Sunyaev \cite{Shakura:1972te} and generalized by Page, Novikov, and Thorne \cite{Novikov:1973kta,Page:1974he}, allow us to infer properties of the central object. Notice that theoretical models of black hole accretion disks have been extensively studied \cite{Abramowicz:2011xu}. Investigating accretion disks of black holes provides insights not only into the central object but also offers a means to test modifications to general relativity. Thin accretion disks and geodesic studies in various background spaces have been well-studied, and their properties hold significant implications for exploring alternative theories \cite{Harko:2008vy,Bambi:2015kza,Guzman:2005bs,Bambi:2013eb,Soroushfar:2016esy,Zhang:2021hit,Liu:2021yev,Kazempour:2017gho,Kazempour:2022asl}.

In this study \cite{Rahaman:2010xs}, the authors introduced a model featuring a black hole surrounded by perfect fluid dark matter. Later, it was revisited in the light of the observed data of Milky Way galaxy \cite{Potapov:2016obe} and studied the effects of dark matter on black hole shadows \cite{Haroon:2018ryd,Hou:2018avu} and circular geodesics\cite{Das:2021otl,Atamurotov:2021hck}. 
Moreover, some researchers have examined the impact of dark matter on accretion disk luminosity. They have investigated various scenarios, including static black holes enveloped by dark matter and black holes surrounded by isotropic/anisotropic pressure dark matter \cite{Boshkayev:2020kle,Kurmanov:2021uqv,Cai:2020kfq}. A recent study \cite{Heydari-Fard:2022xhr} explored the effects of dark matter on the electromagnetic characteristics of slender disks, comparing the findings to those obtained in general relativity for rotating black holes without dark matter. 
In addition, several valuable research in this regard can also be found \cite{Choi:2006gg,Genel:2010pb,Moster:2012fv,Marasco:2021pkl,Xavier:2023exm}.

In this paper, we analyze the accretion disks around the rotating black holes surrounded by two kinds of dark matter halos. We calculate several parameters of the accretion disks of the black holes surrounded by dark matter halos and also compare the ISCOs of the black hole-dark matter spacetimes with Schwarzschild and Kerr black holes.
This paper is organized as follows: In Section \ref{sec:2}, we present and review the spherically symmetric rotating black holes in spacetime surrounded by cold dark matter halo and scalar field dark matter halo, respectively. We also perform an analysis of the horizon of these black holes. In Section \ref{sec:3}, we derive the equations of motion and effective potential. In Section \ref{sec:4}, we calculate all parameters of the accretion disk of rotating black holes surrounded by cold dark matter halo and scalar field dark matter halo. In Section \ref{sec:5}, we present a numerical analysis, including the location of ISCOs for these black holes. Meanwhile, we determine the allowed parameter ranges for the dark matter halo in our models, taking into account the ISCO radius of Sagittarius A*. Finally, in Section \ref{sec:5}, we summarize our findings and conclusions. Throughout the paper, we set $G = c = 1$.
\\

\section{the Black hole dark matter halo spacetimes}\label{sec:2}

In this section, we consider the spherical symmetric rotating black holes in space-time surrounded by two kinds of dark matter halos. We show the space-time metric of both, respectively, derived by Ref. \cite{Xu:2018wow} and we evaluate the horizons of these black holes by considering $g^{rr}=0$.
Notice we represent the horizons of Schwarzschild and Kerr black holes in dimensionless conditions and the complete details of the solutions of these black holes are presented in \cite{Xu:2018wow}. In this paper, we focus on the thin accretion disks of these black holes and some related issues.

\subsection{Cold dark matter (CDM)}\label{CDM}
In this stage, we commence with a rotating black hole surrounded by a cold dark matter halo \cite{Xu:2018wow}.
\begin{widetext}
\begin{eqnarray}\label{ds1}
d s_{1}^{2} = && - \big( 1 - \frac{r^{2}+ 2 M r - r^{2} \big[ 1 +\frac{r}{R_{s}} \big]^{\chi}}{\Sigma^{2}} \big) d t^{2} + \frac{\Sigma^{2}}{\Delta_{1}} d r^{2} + \Sigma^{2} d \theta^{2} + \frac{sin^{2}\theta}{\Sigma^{2}} \big( (r^{2} + a^{2} )^{2} - a^{2} \Delta_{1} sin^{2}\theta \big) d \phi^{2} \nonumber\\ && + \frac{2 \big( r^{2} + 2 M r - r^{2} \big[ 1+\frac{r}{R_{s}} \big]^{\chi} \big) a sin^{2}\theta}{\Sigma^{2}} d\phi dt,
\end{eqnarray}
\end{widetext}
where
\begin{eqnarray}
&&\chi = -\frac{8\pi \rho_{c}R_{s}^{3}}{r}, \nonumber\\
&&\Delta_{1} = r^{2} \big[ 1 + \frac{r}{R_{s}} \big]^{\chi} - 2 M r + a^{2}, \nonumber\\ && \Sigma^{2} = r^{2} + a^{2} cos^{2}\theta .
\end{eqnarray}
\clearpage
\subsection{Scalar field dark matter (SFDM)}\label{SFDM}

Subsequently, we consider the rotating black hole surrounded by a scalar field dark matter halo \cite{Xu:2018wow}.

\begin{widetext}
\begin{eqnarray}\label{ds2}
d s_{2}^{2} = && - \big( 1 - \frac{r^{2}+ 2 M r - r^{2} exp\big[ \varpi \big]}{\Sigma^{2}} \big) d t^{2} + \frac{\Sigma^{2}}{\Delta_{2}} d r^{2} + \Sigma^{2} d \theta^{2} + \frac{sin^{2}\theta}{\Sigma^{2}} \big( (r^{2} + a^{2} )^{2} - a^{2} \Delta_{2} sin^{2}\theta \big) d \phi^{2} \nonumber\\ && + \frac{2 \big( r^{2} + 2 M r - r^{2} exp\big[ \varpi \big] \big) a sin^{2}\theta}{\Sigma^{2}} d\phi dt,
\end{eqnarray}
\end{widetext}
where
\begin{eqnarray}
&&\varpi = -\frac{8 \rho_{c} R^{2}}{\pi}\frac{sin(\frac{\pi r}{R})}{\frac{\pi r}{R}}, \nonumber\\
&&\Delta_{2} = r^{2} exp\big[ \varpi \big] - 2 M r + a^{2},
\end{eqnarray}

\noindent where $\rho_{c}$ is the density of the Universe at the moment when the halo collapsed. 
For CDM, $\rho_{c}/\rho_{NFW}=\frac{r}{R_{s}}(1+\frac{r}{R_{s}})^{2}$ and for SFDM, $\rho_{c}/\rho_{SFDM}=\frac{kr}{sin(kr)}$.
$\rho_{NFW}$ and $\rho_{SFDM}$ are the density profile of CDM and SFDM, respectively, and $k$ is determined by the Compton relationship.
$R_{s}$ and $R$ are the characteristic radius for CDM and SFDM, respectively. 

Furthermore, it should be mentioned that in the considered metrics, the coordinates $r$, $\theta$, and $\phi$ represent standard oblate spheroidal coordinates. The parameters $M$ and $a$ denote the mass and angular momentum of the black holes, respectively. By setting the radii of the dark matter halos, $R_{s}$ for CDM and $R$ for SFDM, to zero, the metrics reduce to the well-known Kerr black hole solution. Notably, the presence of the cross-term 
$dt d\phi$ indicates a coupling between time and motion in the plane of rotation. This coupling vanishes when the black hole's angular momentum approaches zero ($a=0$), resulting in the recovery of the Schwarzschild metric.
In the non-relativistic limit, where the mass $M$ approaches zero, the metric transforms into the orthogonal metric for oblate spheroidal coordinates \cite{Xu:2018wow,Fernandez-Hernandez:2017pgq,Robles:2012uy}.

\begin{figure*}
\centering
{(a) }\includegraphics[width=0.3\linewidth]{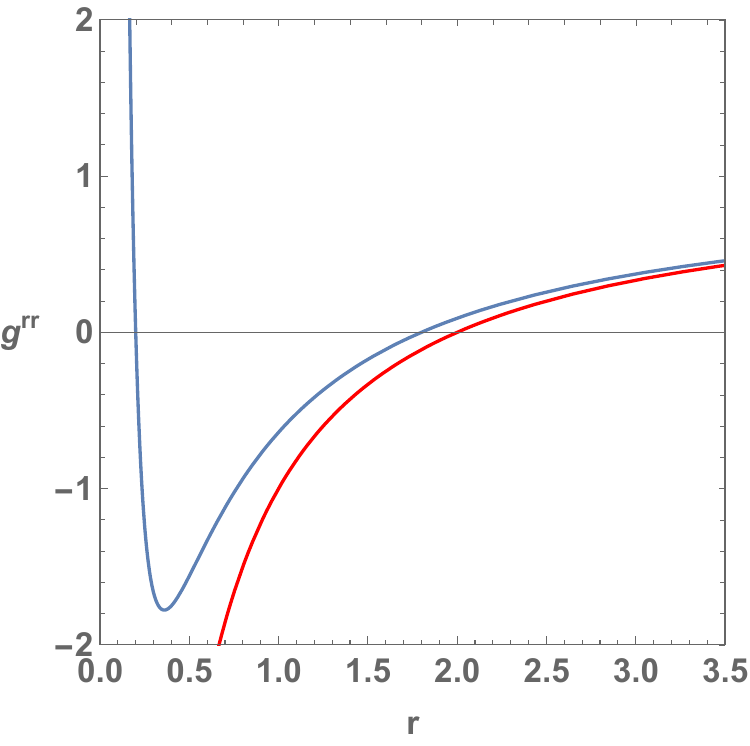}
{(b) }\includegraphics[width=0.3\textwidth]{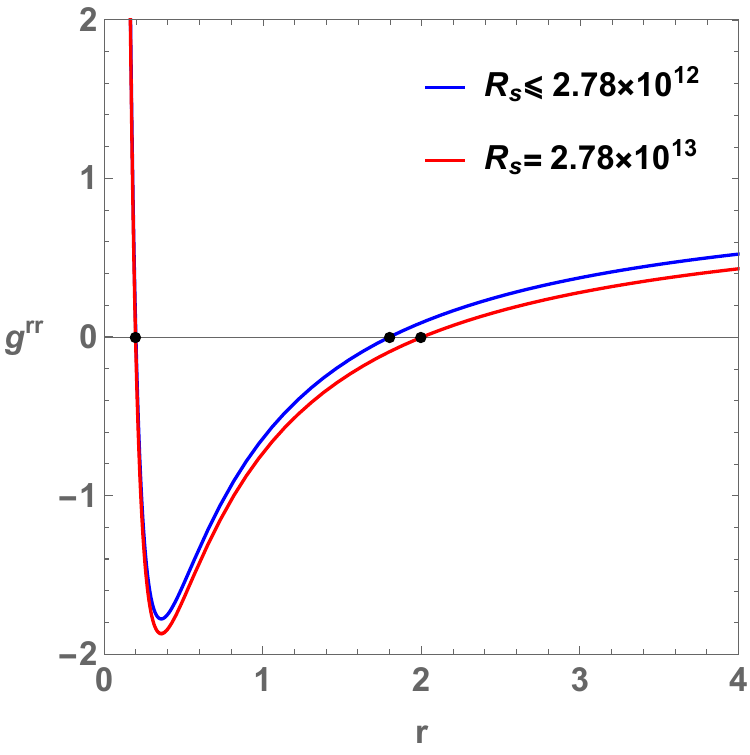}
{(c) }\includegraphics[width=0.3\linewidth]{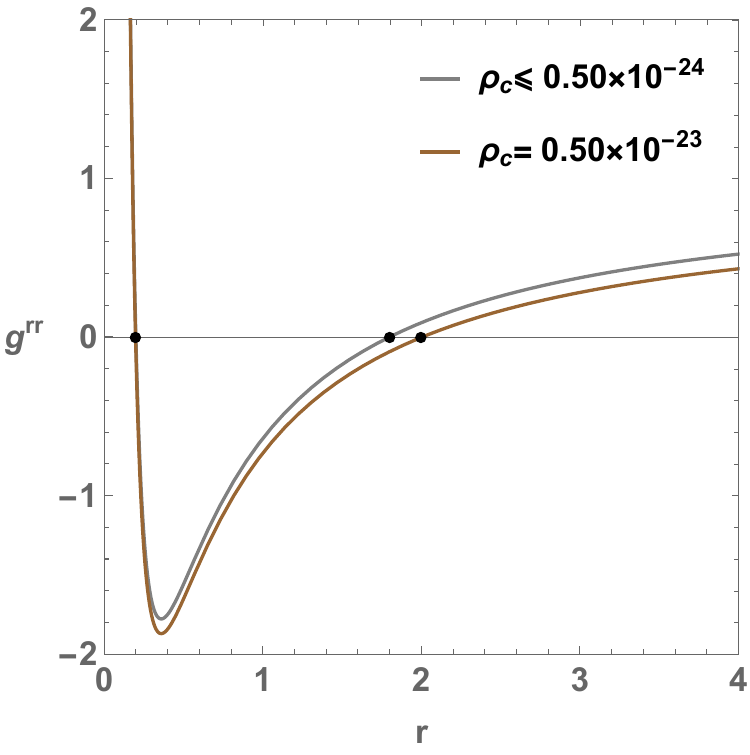}
{(d) }\includegraphics[width=0.3\textwidth]{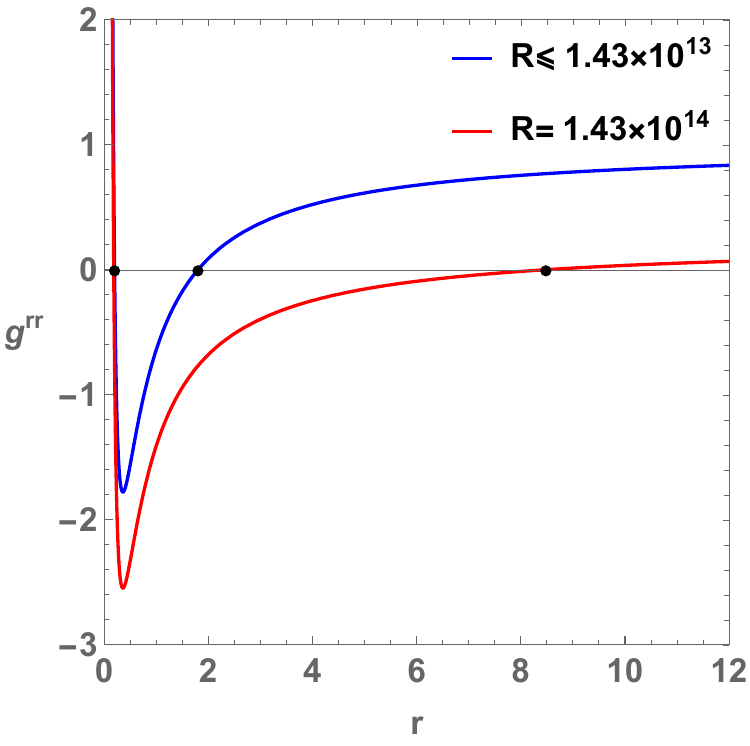}
{(e) }\includegraphics[width=0.3\linewidth]{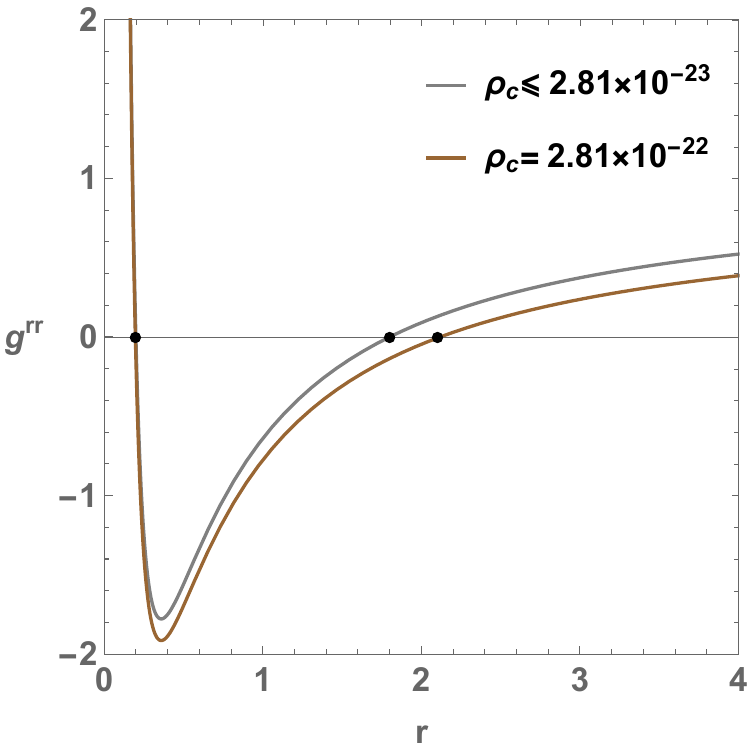}
{(f) }\includegraphics[width=0.3\linewidth]{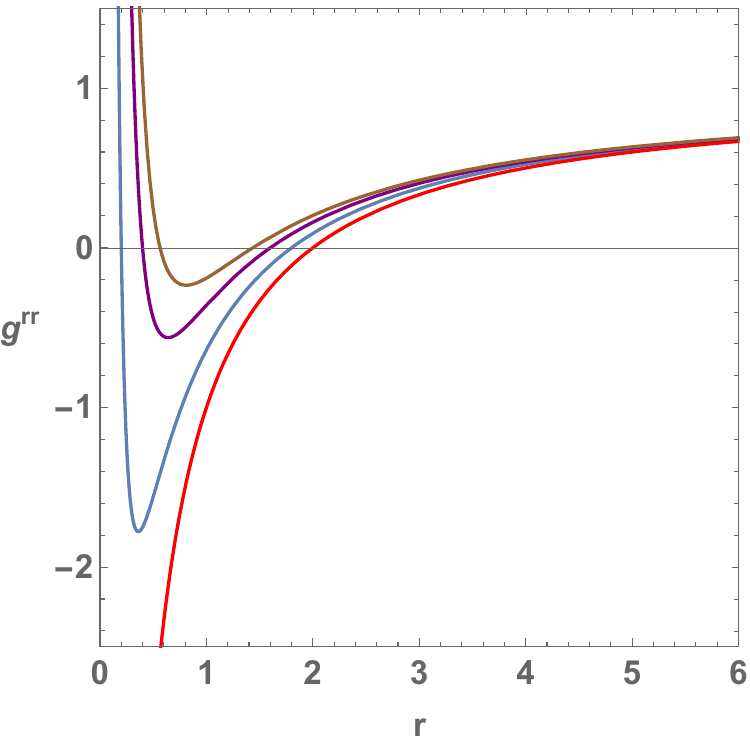}
\caption{The graphs show the profiles of $g^{rr}$ for the black holes. Event horizon radius can be found by solving $g^{rr}=0$ and the number of crossings shows the number of horizons. (a) The red and blue curves show the event horizons of the Schwarzschild ($r_{s}=2$) and Kerr ($r_{-}=0.2$ and $r_{+}=1.8$ by considering $a^{2}=0.36$) black holes in dimensionless conditions, respectively. (b) The graph shows the profile of $g^{rr}$ of Eq. (\ref{ds1}) for the rotating black hole surrounded by a cold dark matter halo for different values of $R_{s}$ with considering $\rho_{c}= 0.50 \times 10^{-29}$ and $a^{2}=0.36$. (c) The graph presents the profile of $g^{rr}$ of Eq. (\ref{ds1}) for the rotating black hole surrounded by a cold dark matter halo for different values of $\rho_{c}$ with considering $R_{s}= 2.78\times 10^{10}$ and $a^{2}=0.36$. (d) The graph shows the profile of $g^{rr}$ of Eq. (\ref{ds2}) for the rotating black hole surrounded by a scalar field dark matter halo for different values of $R$ with considering $\rho_{c}= 2.81 \times 10^{-29}$ and $a^{2}=0.36$. (e) The graph shows the profile of $g^{rr}$ of Eq. (\ref{ds2}) for the rotating black hole surrounded by a scalar field dark matter halo for different values of $\rho_{c}$ with considering $R= 1.43\times 10^{10}$ and $a^{2}=0.36$. (f) The purple curve illustrates the event horizons of the rotating black hole surrounded by a cold dark matter halo. The inner horizon is located $r_{-}=0.4$ and the event horizon is located $r_{+}=1.6$ by considering the values of $a^{2}=0.64$, $R_{s}=2.78 \times 10^{10}$, and $\rho_{c}= 0.50 \times 10^{-29}$. The brown curve illustrates the event horizons of the rotating black hole surrounded by a scalar field dark matter halo. The inner horizon is located $r_{-}=0.56$ and the event horizon is located $r_{+}=1.44$ by considering the values of $a^{2}=0.81$, $R=1.43 \times 10^{10}$, and $\rho_{c}= 2.81 \times 10^{-29}$. The red and blue curves are for the Schwarzschild, and Kerr black holes, respectively.}
\label{fig01}
\end{figure*}

We present the profiles of $g^{rr}$ of those geometries in Figure \ref{fig01}. The figures show the event horizon positions of the black holes for different values of $R_{s}$, $R$, and $\rho_{c}$.

\begin{figure}
\centering
\includegraphics[width=0.4\textwidth]{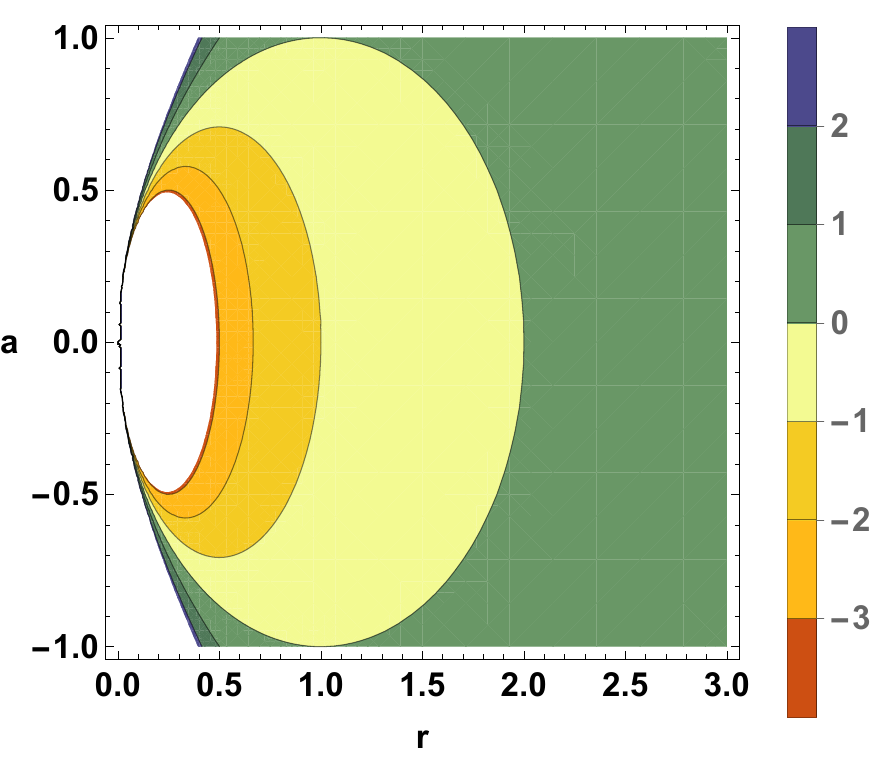}
\caption{The graph shows the profiles of $g^{rr}$ (the event horizons) with respect to different values of $a$ for the Kerr, CDM, and SFDM black holes. Notice that the profiles of $g^{rr}$ (the event horizons) are the same for Kerr, CMD, and SFDM black holes by the values we used for CDM black hole i.e., $R_{s}=2.78\times 10^{10}$, $\rho_{c}= 0.50 \times 10^{-29}$, and for SFDM black hole $R=1.43 \times 10^{10}$, and $\rho_{c}= 2.81 \times 10^{-29}$.}
\label{fig002}
\end{figure}
According to Figs \ref{fig01} and \ref{fig002}, it should be noted that the horizons of CDM and SFDM black holes do not change compared to Kerr black hole by considering the same values of $a$ and using these values i.e., $R_{s}=2.78\times 10^{10}$, $\rho_{c}= 0.50 \times 10^{-29}$, for CDM black hole, and $R=1.43 \times 10^{10}$, $\rho_{c}= 2.81 \times 10^{-29}$ for SFDM black hole.

For plotting the figures \ref{fig01} and \ref{fig002}, we considered $M=1$. In the next sections, we introduced dimensionless parameters to eliminate $M$.
In the following paper, we use the selected values from the Low Surface Brightness galaxy ESO1200211 \cite{Fernandez-Hernandez:2017pgq,Robles:2012uy} which are $R_{s}=5.7kpc$, $\rho_{c}= 0.50 \times 10^{-3}M_{\bigodot}pc^{-3}$ for CDM and $R=2.92kpc$ and $\rho_{c}= 13.66 \times 10^{-3}M_{\bigodot}pc^{-3}$ for SFDM.

\section{Equations of motion}\label{sec:3}

Here, we present the equations of motion and effective potential for evaluating the dynamics of the system. The formation of accretion disks involves particles moving on nearly geodesic orbits, with potential deviations influenced by various physical effects, including electromagnetic forces, pressure gradients, or non-gravitational interactions. In this study, we focus on the idealized scenario of purely geodesic motion, neglecting these additional factors for simplicity.
In the equatorial plane $\theta = \frac{\pi}{2}$, we have two constants of motion which are the conserved energy $E$ and angular momentum $L$. 
\begin{eqnarray}\label{E}
g_{tt}\dot{t} + g_{t\phi}\dot{\phi} = - E,
\end{eqnarray}

\begin{eqnarray}\label{L}
g_{t\phi}\dot{t} + g_{\phi\phi}\dot{\phi} = L,
\end{eqnarray}
Note that a dot represents a derivative with respect to the affine parameter $\tau$. Indeed, $\tau$ serves as an affine parameter along the geodesic, and we specifically choose it to represent the proper time. The relationship between $\tau$ and the coordinate time is given by $d \tau = \sqrt{g_{00}} dt$ \cite{Landau:1975pou}. According to Eq. (\ref{E}) and Eq. (\ref{L}), we can find $t$ and $\phi$ components of the 4-velocity $\dot{x}^{\mu}$ as
\begin{eqnarray}
\frac{d t}{d \tau} = \frac{E g_{\phi\phi} + L g_{t\phi}}{g_{t\phi}^{2} - g_{tt}g_{\phi\phi}},
\end{eqnarray}

\begin{eqnarray}
\frac{d \phi}{d \tau} = - \frac{E g_{t\phi} + L g_{tt}}{g_{t\phi}^{2} - g_{tt}g_{\phi\phi}}.
\end{eqnarray}
Notice that from the normalization condition, $g_{\mu\nu}\dot{x}^{\mu}\dot{x}_{\nu} = -1$, we have
\begin{eqnarray}
V_{eff}(r, \theta) = g_{rr}\dot{r}^{2} + g_{\theta\theta}\dot{\theta}^{2},
\end{eqnarray}
so, the effective potential is given,
\begin{eqnarray}
V(r, \theta) = -1 + \frac{E^{2}g_{\phi\phi} + 2 E L g_{t\phi} + L^{2}g_{tt}}{g_{t\phi}^{2} - g_{tt}g_{\phi\phi}}.
\end{eqnarray}

Here, we present the dimensionless quantities from now on.
\begin{eqnarray}\label{Dm}
\tilde{r} = \frac{r}{M}, \qquad \tilde{L} = \frac{L}{M}, \qquad \tilde{a}= \frac{a}{M}, \nonumber\\ \tilde{R}_{s} = \frac{R_{s}}{M}, \qquad \tilde{R} = \frac{R}{M}, \qquad \tilde{\rho_{c}}= \rho_{c}M^{2}.
\end{eqnarray}

It's worth noting that in terms of unit conversion, $\rho_{c} = \tilde{\rho}_{c} \times (4.86 \times 10^{26})M_{\bigodot}p c^{-3}$ and $R_{s} = \tilde{R}_{s} \times (2.048 \times 10^{-10}) kpc$. In the following, we present the effective potential and the $t$ and $\phi$ components of the 4-velocity $\dot{x}^{\mu}$ for CDM and SFDM black holes. We have calculated all these parameters in dimensionless conditions which we introduced in Eq. (\ref{Dm}). 

\subsection{Cold dark matter (CDM)}\label{subsec31}

\begin{widetext}
\begin{eqnarray}
(\frac{d t}{d \tau})_{1} = \frac{\big[ \tilde{a}\tilde{L} (2 + \tilde{r}) + \tilde{r}^{3} E + 2 \tilde{a}^{2} (1 + \tilde{r}) E \big] (1 + \frac{\tilde{r}}{\tilde{R}_{s}})^{\tilde{\chi}} - \tilde{a}\tilde{r}(\tilde{L} + \tilde{a}E)}{\tilde{r}^{3} + \tilde{r} (\tilde{a}^{2} - 2 \tilde{r})(1 + \frac{\tilde{r}}{\tilde{R}_{s}})^{\tilde{\chi}}}, 
\end{eqnarray}
\begin{eqnarray}
(\frac{d \phi}{d \tau})_{1} = \frac{\tilde{r}(\tilde{L} + \tilde{a}E) - \big[ 2 \tilde{L} + \tilde{a}(2 + \tilde{r}) E \big] (1 + \frac{\tilde{r}}{\tilde{R}_{s}})^{\tilde{\chi}}}{\tilde{r}^{3} + \tilde{r}(\tilde{a}^{2} - 2 \tilde{r})(1 + \frac{\tilde{r}}{\tilde{R}_{s}})^{\tilde{\chi}}},
\end{eqnarray}

\begin{eqnarray}\label{V1}
\tilde{V}(eff)_{1} = \frac{\big[ 2 \tilde{L}^{2} + 2 \tilde{a}\tilde{L}(2 + \tilde{r}) E + \tilde{r}^{2}(2 + \tilde{r}E^{2}) +\tilde{a}^{2} ( 2(1+\tilde{r})E^{2} - \tilde{r}) \big] (1 + \frac{\tilde{r}}{\tilde{R}_{s}})^{\chi} - \tilde{r}\big( \tilde{r}^{2} + (\tilde{L} + \tilde{a}E)^{2}\big)}{\tilde{r}^{3} + \tilde{r}(\tilde{a}^{2} - 2 \tilde{r})(1 + \frac{\tilde{r}}{\tilde{R}_{s}})^{\tilde{\chi}}}.
\end{eqnarray}
\end{widetext}

\subsection{Scalar field dark matter (SFDM)}\label{subsec32}

\begin{widetext}
\begin{eqnarray}
(\frac{d t}{d \tau})_{2} = \frac{\big[ \tilde{a}\tilde{L} (2 + \tilde{r}) + \tilde{r}^{3}E + 2 \tilde{a}^{2}(1 + \tilde{r})E \big] exp[\tilde{\varpi}] - \tilde{a}\tilde{r}(\tilde{L} + \tilde{a}E)}{exp[\tilde{\varpi}](\tilde{a}^{2} - 2 \tilde{r})\tilde{r} + \tilde{r}^{3}},
\end{eqnarray}
\begin{eqnarray}
(\frac{d \phi}{d \tau})_{2} = \frac{\tilde{r}(\tilde{L}+\tilde{a}E) - exp[\tilde{\varpi}]\big[ 2 \tilde{L} + \tilde{a}(2 + \tilde{r})E \big]}{exp[\tilde{\varpi}](\tilde{a}^{2} - 2 \tilde{r})\tilde{r} + \tilde{r}^{3}},
\end{eqnarray}

\begin{eqnarray}\label{V2}
\tilde{V}(eff)_{2} = \frac{exp[\tilde{\varpi}]\big[ 2 \tilde{L}^{2} - \tilde{a}^{2}\tilde{r} + 2 \tilde{r}^{2} + 2 \tilde{a}\tilde{L}(2+ \tilde{r})E + ( \tilde{r}^{3} + 2 \tilde{a}^{2}(1 + \tilde{r})) E^{2} \big] - \tilde{r} \big( \tilde{r}^{2}+(\tilde{L} + \tilde{a}E)^{2}\big)}{exp[\tilde{\varpi}](\tilde{a}^{2} - 2 \tilde{r})\tilde{r} + \tilde{r}^{3}}.
\end{eqnarray}
\end{widetext}

\section{Thin accretion disks}\label{sec:4}

In this stage, we can calculate all parameters of the thin accretion disks of black holes.
Therefore, the specific energy $\tilde{E}$, the specific angular momentum $\tilde{L}$, the angular velocity $\Omega$ and the flux of the radiant energy $F$, over the disk of the particles which move in circular orbits, can be obtained.
The physical properties of the accretion disk can be derived from fundamental structure equations that ensure the conservation of mass, energy, and angular momentum. Notably, the kinematic quantities depend on the orbital radius and can be determined using general expressions presented in references \cite{Novikov:1973kta,Page:1974he,Thorne:1974ve}. Additionally, we utilize dimensionless quantities, as defined in Equation (\ref{Dm}), to calculate all relevant parameters.
\begin{eqnarray}
\tilde{L} = \frac{g_{t\phi} + g_{\phi\phi}\Omega}{\sqrt{-g_{tt} - 2 g_{t\phi}\Omega - g_{\phi\phi}\Omega^{2}}},
\end{eqnarray}
\begin{eqnarray}
\tilde{E} = - \frac{g_{tt} + g_{t\phi}\Omega}{\sqrt{-g_{tt} - 2 g_{t\phi}\Omega - g_{\phi\phi}\Omega^{2}}},
\end{eqnarray}
\begin{eqnarray}
\Omega = \frac{- g_{t\phi , r} + \sqrt{(g_{t\phi , r})^{2} - g_{tt, r}g_{\phi\phi, r}}}{g_{\phi\phi, r}}.
\end{eqnarray}
Furthermore, the radiant energy flux emanating from the accretion disk can be calculated by applying the conservation equations of rest mass, energy, and the angular momentum of the disk particles.
\begin{eqnarray}
F (r)=\frac{-\dot{M}_{0}}{4\pi\sqrt{-g}}\frac{\Omega_{,r}}{(\tilde{E}-\Omega \tilde{L})^{2}}\int_{r_{ISCO}}^{r}(\tilde{E}-\Omega \tilde{L})\tilde{L}_{,r}dr, \nonumber\\
\end{eqnarray}
where the $\dot{M}_{0}$ is the mass accretion rate.

\subsection{Cold dark matter (CDM)}

We have obtained the specific angular momentum $\tilde{L}_{1}$, the specific energy $\tilde{E}_{1}$, and the angular velocity $\Omega_{1}$ of the thin accretion disk of the rotating black hole surrounded by a cold dark matter halo in dimensionless conditions, as below;
\begin{widetext}
\begin{eqnarray}
\tilde{L}_{1} = \frac{\bigg[ -\tilde{a}\tilde{r} (1 + \tilde{a} \Omega_{1}) + \big( \tilde{a}( 2 + \tilde{r}) + \tilde{r}^{3}\Omega_{1}+ 2 \tilde{a}^{2}(1 + \tilde{r})\Omega_{1} \big) (1+\frac{\tilde{r}}{\tilde{R}_{s}})^{\tilde{\chi}} \bigg] (1+ \frac{\tilde{r}}{\tilde{R}_{s}})^{\tilde{\chi}}}{\tilde{r}\bigg[ -\big( 2 + 2 \Omega_{1} \tilde{a}(2 + \tilde{r}) + \big[ \tilde{r}^{3} + 2 \tilde{a}^{2} ( 1 + \tilde{r}) \big]\Omega_{1}^{2}\big) \tilde{r}^{-1} + (1 + \tilde{a}\Omega_{1})^{2} (1 + \frac{\tilde{r}}{\tilde{R}_{s}})^{\tilde{\chi}} \bigg]^{\frac{1}{2}}},
\end{eqnarray}
\begin{eqnarray}
\tilde{E}_{1} = \frac{\bigg[ \tilde{r} + \tilde{a}\tilde{r}\Omega_{1} - \big( \Omega_{1} \tilde{a}(\tilde{r} + 2 ) + 2 \big) (1+\frac{\tilde{r}}{\tilde{R}_{s}})^{\tilde{\chi}} \bigg] (1+ \frac{\tilde{r}}{\tilde{R}_{s}})^{\tilde{\chi}}}{\tilde{r}\bigg[ -\big( 2 + 2 \Omega_{1} \tilde{a}(2 + \tilde{r}) + \big[ \tilde{r}^{3} + 2 \tilde{a}^{2} ( 1 + \tilde{r}) \big] \Omega_{1}^{2} \big) \tilde{r}^{-1} + (1 + \tilde{a}\Omega_{1})^{2} (1 + \frac{\tilde{r}}{\tilde{R}_{s}})^{\tilde{\chi}} \bigg]^{\frac{1}{2}}},
\end{eqnarray}
\begin{eqnarray}
\Omega_{1} = - \Bigg\lbrace 4 \tilde{a} \pi \tilde{r} \big( -1 + Log[1 + \frac{\tilde{r}}{\tilde{R}_{s}}] \big) \tilde{R}_{s}^{3} \tilde{\rho}_{c} + 4 \tilde{a} \pi Log[1 + \frac{\tilde{r}}{\tilde{R}_{s}}] \tilde{R}_{s}^{4} \tilde{\rho}_{c} + (\tilde{r} + \tilde{R}_{s}) (1 + \frac{\tilde{r}}{\tilde{R}_{s}})^{-\tilde{\chi}} \bigg[ \tilde{a} \nonumber\\ + \tilde{r} \bigg( \tilde{r} + 4 \pi \tilde{r} (1 + \frac{\tilde{r}}{\tilde{R}_{s}})^{\tilde{\chi}}\tilde{R}_{s}^{3}\tilde{\rho}_{c} \big[ \tilde{r} (Log[1+ \frac{\tilde{r}}{\tilde{R}_{s}}] - 1 ) + Log[ 1 + \frac{\tilde{r}}{\tilde{R}_{s}} ] \tilde{R}_{s} \big] (\tilde{r} + \tilde{R}_{s} )^{-1} \bigg)^{\frac{1}{2}} \bigg] \Bigg\rbrace \nonumber\\ \Bigg[ (\tilde{a}^{2} - \tilde{r}^{3}) (1 + \frac{\tilde{r}}{\tilde{R}_{s}})^{-\tilde{\chi}} (\tilde{r} +\tilde{R}_{s}) + 4 \tilde{a}^{2}\tilde{R}_{s}^{3}\pi \big( \tilde{r}(Log[1+ \frac{\tilde{r}}{\tilde{R}_{s}}] - 1) + Log[1+\frac{\tilde{r}}{\tilde{R}_{s}}] \tilde{R}_{s} \big) \tilde{\rho}_{c} \Bigg]^{-1}.
\end{eqnarray}
\end{widetext}
\begin{figure*}
\centering
{(a) }\includegraphics[width=0.29\linewidth]{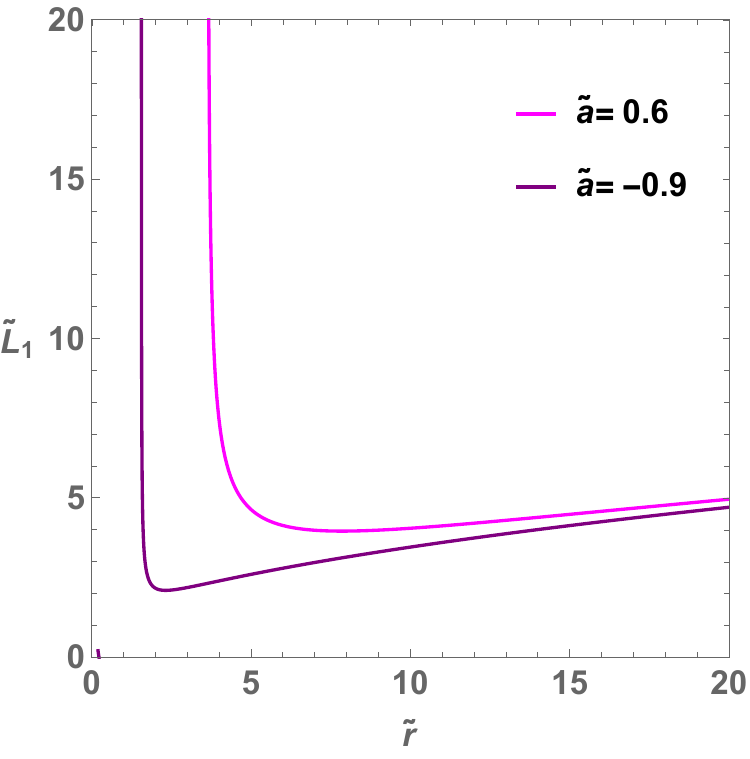}
{(b) }\includegraphics[width=0.285\linewidth]{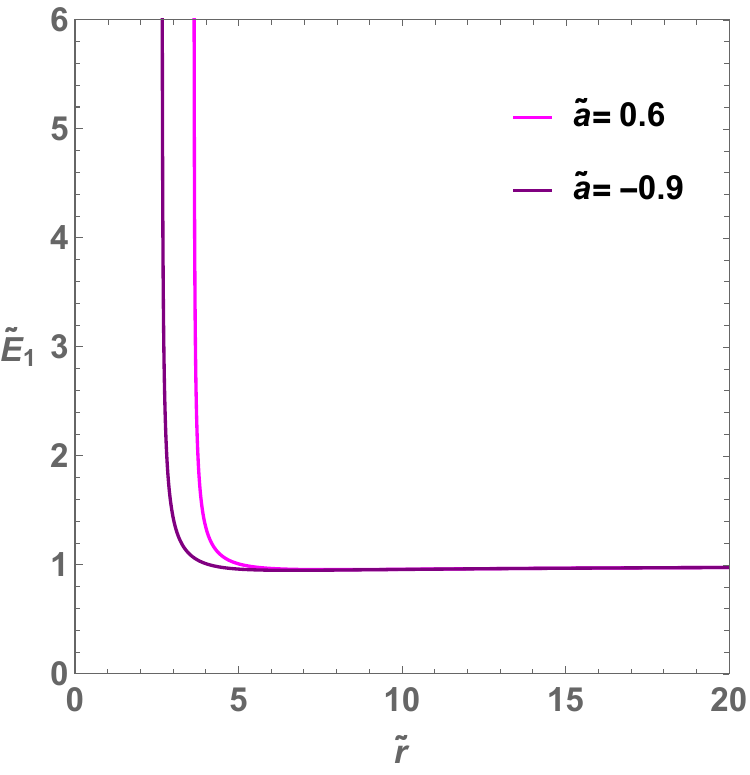}
{(c) }\includegraphics[width=0.3\textwidth]{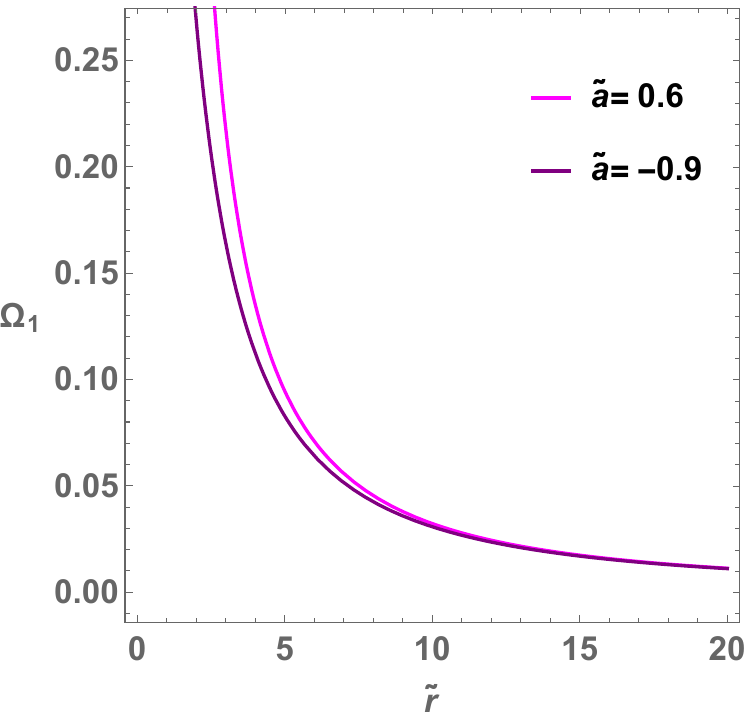}
\caption{The graphs depict the changing parameters of the thin accretion disks of a rotating black hole surrounded by a cold dark matter halo, by considering the values of $\tilde{R}_{s}=2.78\times 10^{10}$, and $\tilde{\rho}_{c}= 0.50 \times 10^{-29}$. (a) Changes of the specific angular momentum $\tilde{L}_{1}$. (b) Changes of the specific energy $\tilde{E}_{1}$. (c) Changes in the angular velocity $\Omega_{1}$.}
\label{fig44}
\end{figure*}
In Fig. \ref{fig44}, the specific angular momentum $\tilde{L}_{1}$, the specific energy $\tilde{E}_{1}$ and the angular velocity $\Omega_{1}$ of the CDM rotating black hole of the thin accretion disk versus $\tilde{r}$ are plotted for two values of $\tilde{a}=-0.9$ (retrograde) and $\tilde{a}=0.6$ (direct).

From an overall perspective, the most striking feature of Fig. \ref{fig44}. (a) is that the specific angular momentum $\tilde{L}_{1}$ witnessed three phases. In the first phase, the specific angular momentum increased sharply between $\tilde{r}=2$ and $\tilde{r}=5$ towards small r. The second phase is the lowest level of specific angular momentum which is around $\tilde{r}=3$ for retrograde i.e., $\tilde{a}=-0.9$, and around $\tilde{r}=7$ for direct i.e., $\tilde{a}=0.6$. In the third phase, the specific angular momentum increases gradually for both cases.
Regarding the specific energy $\tilde{E}_{1}$ in Fig. \ref{fig44} (b), it should be mentioned that they increase sharply between $\tilde{r}=3$ and $\tilde{r}=4$ towards small r for both cases of retrograde and direct. Then they remain unchanged.
Concerning the angular velocity $\Omega_{1}$ in Fig. \ref{fig44} (c), it is obvious that the graphs start to grow from the location of $\tilde{r}=10$ sharply towards small r. In the following, the increasing trend continues strongly for both cases $\tilde{a}=-0.9$ and $\tilde{a}=0.6$ near the event horizons.

\subsection{Scalar field dark matter (SFDM)}

We have determined the specific angular momentum $\tilde{L}_{2}$, the specific energy $\tilde{E}_{2}$, and the angular velocity $\Omega_{2}$ of the thin accretion disk of the rotating black hole surrounded in a scalar field dark matter halo under dimensionless conditions as below;
\begin{widetext}
\begin{eqnarray}
\tilde{L}_{2} = \frac{\tilde{a}\tilde{r}^{-2}\bigg( 2 \tilde{r} + \tilde{r}^{2} - \tilde{r}^{2} exp[\tilde{\varpi}] \bigg) + \bigg[ 2 \tilde{a}^{2}\big( 1 + \frac{1}{\tilde{r}} \big) + \tilde{r}^{2} - \tilde{a}^{2} exp[\tilde{\varpi}] \bigg]\Omega_{2}}{\bigg[ 1 - \big( 2 \tilde{r} + \tilde{r}^{2} - \tilde{r}^{2} exp[\tilde{\varpi}] \big) \tilde{r}^{-2} - 2 \tilde{a}\Omega_{2} \big( 2 \tilde{r} + \tilde{r}^{2} - \tilde{r}^{2} exp[\tilde{\varpi}] \big) \tilde{r}^{-2} - \big( 2 \tilde{a}^{2} (1+\frac{1}{\tilde{r}}) + \tilde{r}^{2} - \tilde{a}^{2}exp[\tilde{\varpi}] \big) \Omega_{2}^{2} \bigg]^{\frac{1}{2}}},
\end{eqnarray}
\begin{eqnarray}
\tilde{E}_{2} = - \frac{-1 + \big( 2 \tilde{r} + \tilde{r}^{2} - \tilde{r}^{2} exp[\tilde{\varpi}] \big) \tilde{r}^{-2} + \tilde{a}\Omega_{2} \big( 2 \tilde{r} + \tilde{r}^{2} - \tilde{r}^{2}exp[\tilde{\varpi}] \big) \tilde{r}^{-2}}{\bigg[ 1 - \big( 2 \tilde{r} + \tilde{r}^{2} - \tilde{r}^{2} exp[\tilde{\varpi}] \big) \tilde{r}^{-2} - 2 \tilde{a}\Omega_{2} \big( 2 \tilde{r} + \tilde{r}^{2} - \tilde{r}^{2} exp[\tilde{\varpi}] \big) \tilde{r}^{-2} - \big( 2 \tilde{a}^{2} (1+\frac{1}{\tilde{r}}) + \tilde{r}^{2} - \tilde{a}^{2}exp[\tilde{\varpi}] \big) \Omega_{2}^{2} \bigg]^{\frac{1}{2}}},
\end{eqnarray}
\begin{eqnarray}
\Omega_{2} = \frac{4 \tilde{a}\tilde{R}^{2}\big( \tilde{R} sin[\frac{\pi \tilde{r}}{\tilde{R}}] - \pi \tilde{r} cos[\frac{\pi \tilde{r}}{\tilde{R}}] \big) \tilde{\rho}_{c} + \pi exp[- \tilde{\varpi}] \bigg[ \tilde{a} \pi + \tilde{r} \big( \tilde{r}\pi^{2} + 4 \tilde{r}\tilde{R}^{2}\tilde{\rho}_{c} ( \tilde{R} sin[\frac{\pi \tilde{r}}{\tilde{R}}] - \pi \tilde{r} cos[\frac{\pi \tilde{r}}{\tilde{R}}]) \big)^{\frac{1}{2}} \bigg]}{exp[-\tilde{\varpi}] \pi^{2}(\tilde{r}^{3}- \tilde{a}^{2}) + 4 \tilde{a}^{2}\tilde{R}^{2} \big( \pi \tilde{r} cos[\frac{\pi \tilde{r}}{\tilde{R}}] - \tilde{R} sin[\frac{\pi \tilde{r}}{\tilde{R}}] \big) \tilde{\rho}_{c}}.
\end{eqnarray}
\end{widetext}

\begin{figure*}
\centering
{(a) }\includegraphics[width=0.29\linewidth]{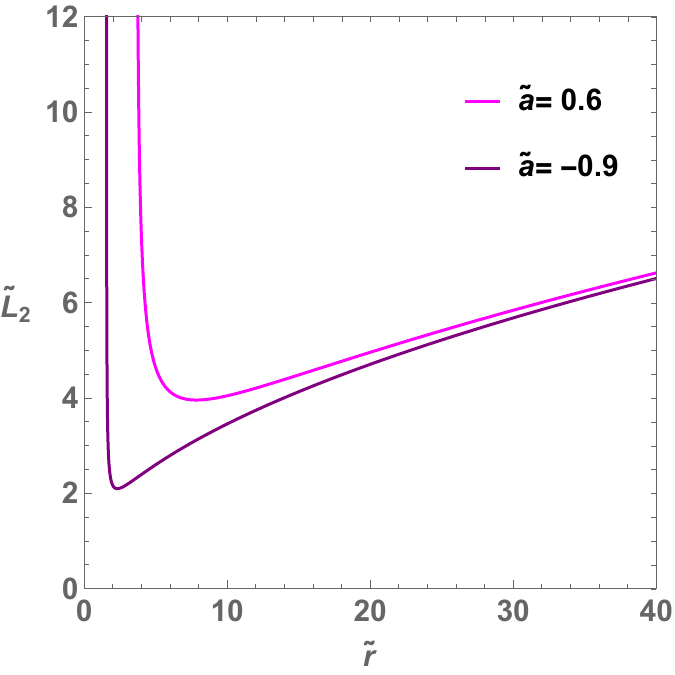}
{(b) }\includegraphics[width=0.3\linewidth]{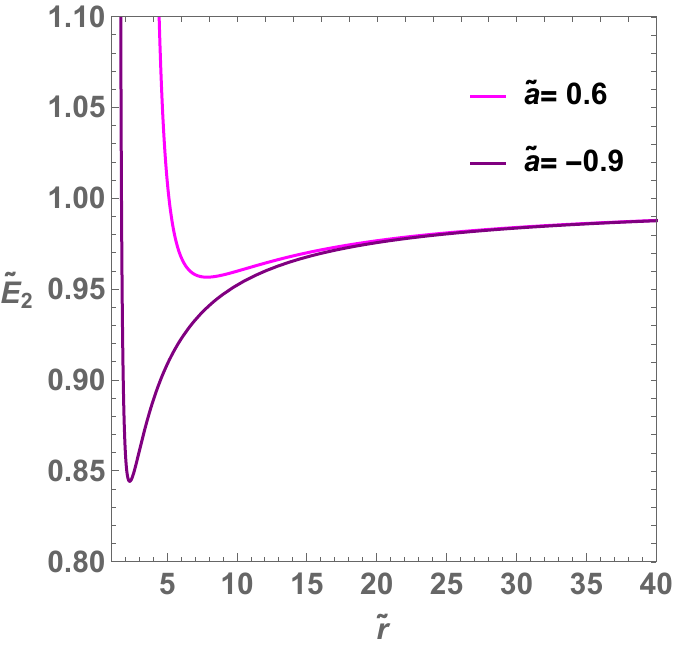}
{(c) }\includegraphics[width=0.3\textwidth]{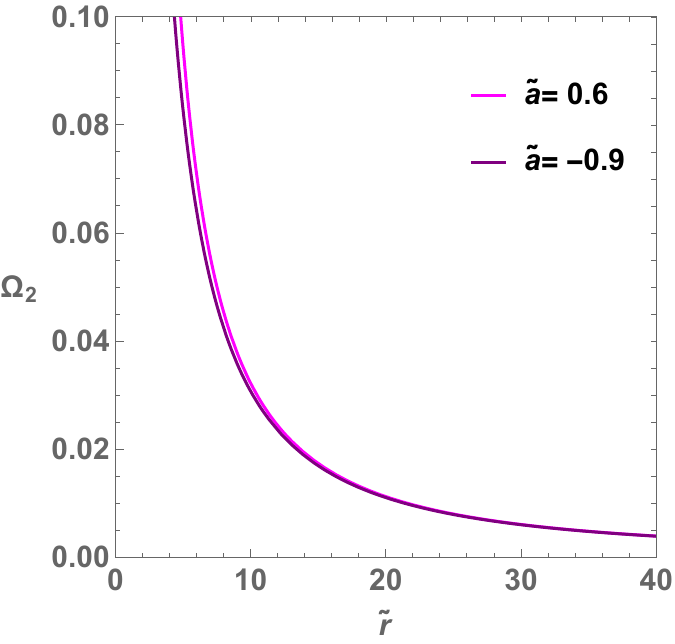}
\caption{The graphs depict the changing parameters of the thin accretion disks of a rotating black hole surrounded by a scalar field dark matter halo, by considering the values of $\tilde{ R}=1.43 \times 10^{10}$, and $\tilde{\rho}_{c}= 2.81 \times 10^{-29}$. (a) Changes of the specific angular momentum $\tilde{L}_{2}$ (b) Changes of the specific energy $\tilde{E}_{2}$. (c) Changes in the angular velocity $\Omega_{2}$.}
\label{fig45}
\end{figure*}

In Fig. \ref{fig45}, the specific angular momentum $\tilde{L}_{2}$, the specific energy $\tilde{E}_{2}$ and the angular velocity $\Omega_{2}$ of the SFDM rotating black hole of the thin accretion disk versus $\tilde{r}$ are illustrated for two values of $\tilde{a}=-0.9$ (retrograde) and $\tilde{a}=0.6$ (direct). 

According to Fig. \ref{fig45}. (a), one can immediately find out that the specific angular momentum $\tilde{L}_{2}$ rises for both cases of $\tilde{a}$ towards small values of $\tilde{r}$ around $\tilde{r}=2$ and $\tilde{r}=4$ for retrograde and direct motion, respectively. Moreover, the graphs reached the minimums which are approximately $\tilde{r}=3$ and $\tilde{r}=7$. In the larger values of $\tilde{r}$, the angular momentum $\tilde{L}_{2}$ increases.
Furthermore, Fig. \ref{fig45}. (b) shows the specific energy $\tilde{E}_{2}$ for both cases. Specific energies increase mildly at long distances. But as they approach the black hole, they reach a minimum value around $\tilde{r}=2$ and $\tilde{r}=7$ for retrograde and direct cases, respectively. Then they reach their maximum values near the horizons.
Finally, Fig. \ref{fig45}. (c) presents the changes of the specific angular momentum $\Omega_{2}$. The specific angular momentum rises towards the horizon.

It is worth pointing out that at the ISCOs, the specific energy and angular momentum reach their minimum values for stable orbits.

While the behavior of the specific angular momentum in both the CDM and SFDM cases exhibits similarities, there are notable differences in the specific energy and angular velocity as $r$ varies. 
Regarding the specific energy in the case of CDM, it's worth noting that the minimum values for both retrograde and direct motions are found around $r=3$ to $r=4$, whereas in the SFDM case, they are located around $r=2$ to $r=7$.

Additionally, it is intriguing to highlight the subtle differences in the angular velocity between the two cases. Although the general forms are comparable, the locations of the maximum values differ. In the CDM case, the maximum values of the angular velocity lie within the range of $r=2$ to $r=3$, while in the SFDM case, they are found within the range of $r=5$ to $r=6$.

\section{Numerical Analysis}\label{sec:5}

Now we would like to focus on the ISCOs. Notice that the ISCOs can be obtained by setting to zero the second derivative of the effective potential. Therefore, the dimensionless radius of the innermost stable circular geodesic orbits can be obtained.

It is worth pointing out that to have circular orbits, we should have $\tilde{V}_{eff}=0$, and $\frac{d \tilde{V}_{eff}}{d r}=0$. 
Furthermore, innermost circular orbits occur at the local minimum of the effective potential, thus the innermost (marginally) stable circular geodesic orbit is obtained from $\frac{d^{2} \tilde{V}_{eff}}{d r^{2}} = 0$. Also, it is obvious that the ISCO is the transition between stable and unstable circular orbits \cite{Novikov:1973kta,Jiang:2021ajk,Kazempour:2022asl}.

\begin{figure}
\centering
\includegraphics[width=7cm]{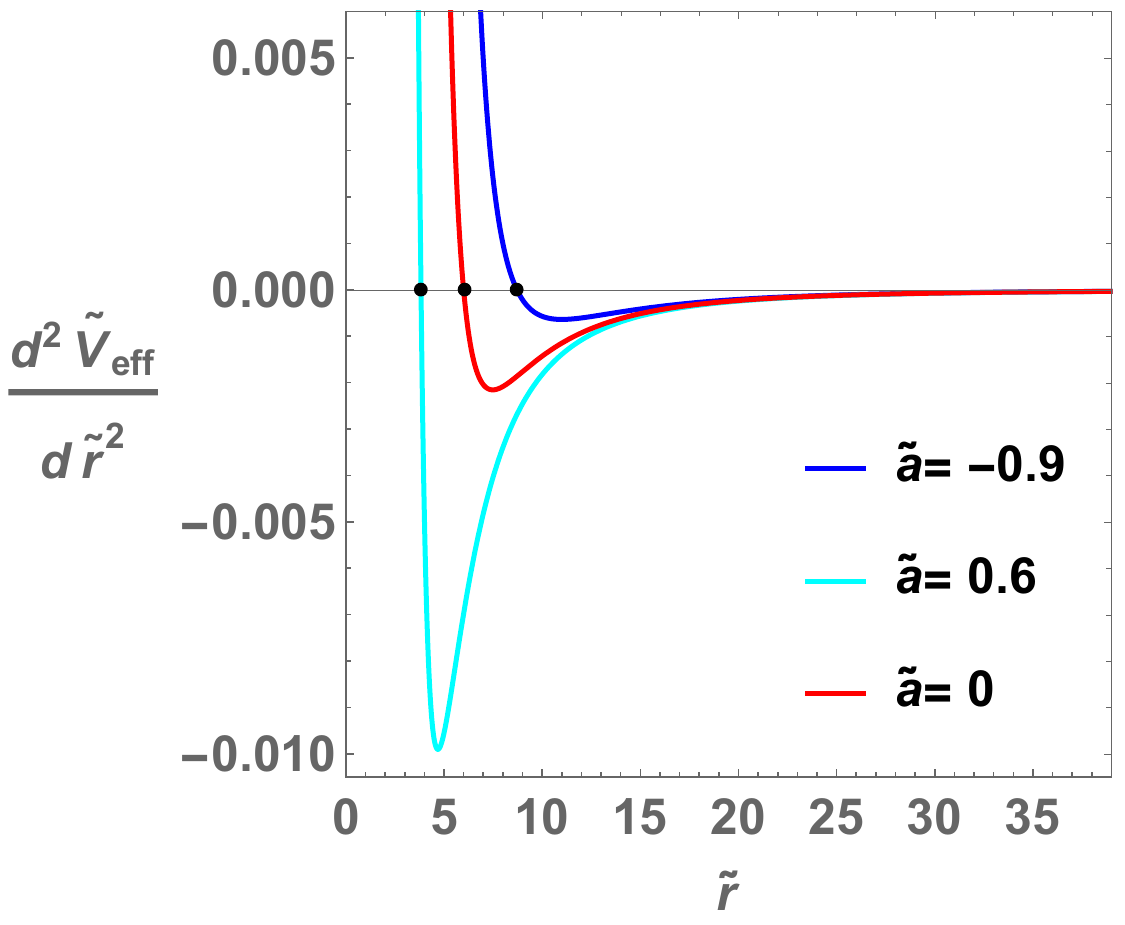}
\caption[figs]
{The second derivative of the effective potential of Kerr and Schwarzschild black holes shows the locations of the ISCOs for the cases, $\tilde{a}=-0.9$ (Retrograde), $\tilde{a}=0.6$ (Direct), and $\tilde{a}=0$.}
\label{figSDEPKKS}
\end{figure}

In fig. \ref{figSDEPKKS}, we explore the concept of ISCOs by illustrating the second derivative of the effective potential of Kerr and Schwarzschild black holes with respect to $\tilde{r}$, for three different cases: retrograde motion with $\tilde{a}=-0.9$, direct motion with $\tilde{a}=0.6$, and non-rotating black holes with $\tilde{a}=0$. The figure illustrates the connection between the second derivative of the effective potential and the ISCOs, which are characterized by the solutions to the equation $\frac{d^{2}\tilde{V}_{eff}}{d\tilde{r}^{2}}=0$. Our study reveals that the ISCOs for retrograde motion occur at $\tilde{r}_{ISCO}=8.75$, while those for direct motion occur at $\tilde{r}_{ISCO}=3.83$. Schwarzschild black hole has ISCO at $\tilde{r}_{ISCO}=6$. 
To show the validity of the method, we solely derived the standard ISCOs for Kerr and Schwarzschild black holes in dimensionless form, which are completely consistent with the results \cite{Bardeen:1972fi,Carroll:1997ar}.

\begin{table*}
\caption{\label{tab:table1}Locations of the event horizons and the ISCOs in dimensionless conditions by considering the $\tilde{a}=-0.9$, and $\tilde{a}=0.6$ (retrograde orbit and direct orbit, respectively) for Schwarzschild black hole, Kerr black hole and CDM, and SFM black holes by considering the values of $\tilde{R_{s}}=2.78 \times 10^{10}$, $\tilde{\rho_{c}}= 0.50 \times 10^{-29}$ for a CDM and $\tilde{R}=1.43 \times 10^{10}$ and $\tilde{\rho_{c}}= 2.81 \times 10^{-29}$ for SFDM.}
\begin{tabular}{cccccc}\hline
Black Hole &Inner horizon $\tilde{r}_{-}$&Event horizon $\tilde{r}_{+}$&Innermost Stable circular orbits \\ \hline
Schwarzschild ($\tilde{a}=0$)& - & 2 & 6 \\ \hline
Kerr (retrograde, $\tilde{a}=-0.9$)&0.56&1.44& 8.75 \\ \hline
Kerr (retrograde, $\tilde{a}=-1$)&-&1& 9 \\ \hline
Kerr (direct, $\tilde{a}=0.6$)&0.2&1.8 & 3.83 \\ \hline
CDM (retrograde, $\tilde{a}=-0.9$)&0.56&1.44&$2.32$\\ \hline
CDM (direct, $\tilde{a}=0.6$)&0.2&1.8 &7.85\\ \hline
SFDM (retrograde, $\tilde{a}=-0.9$)&0.56&1.44&$2.32$\\ \hline
SFDM (direct, $\tilde{a}=0.6$)&0.2&1.8&7.85\\ \hline
\end{tabular}
\end{table*}

\begin{figure*}
\centering
{(a)}\includegraphics[width=5cm]{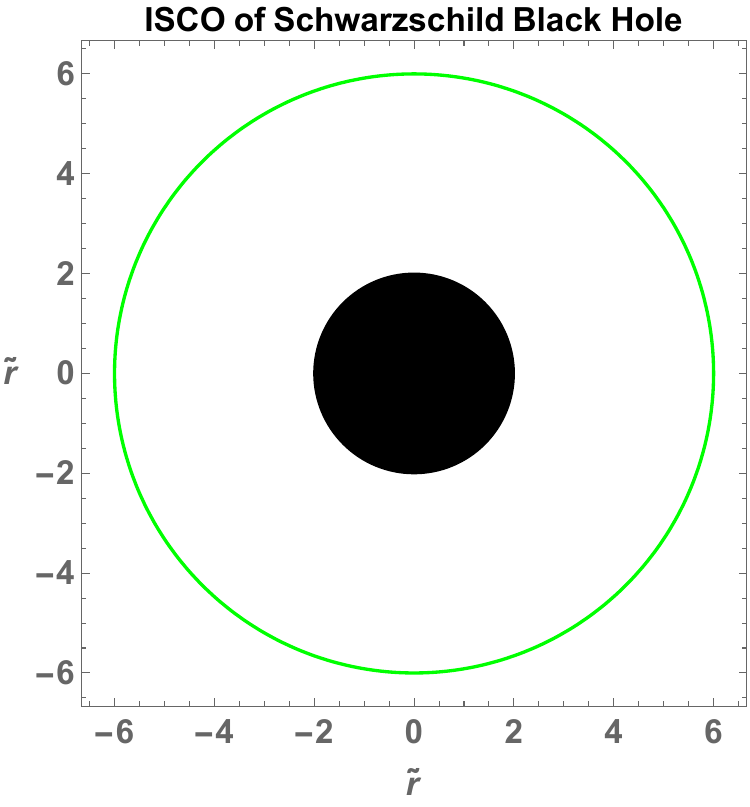}
{(b)}\includegraphics[width=5.15cm]{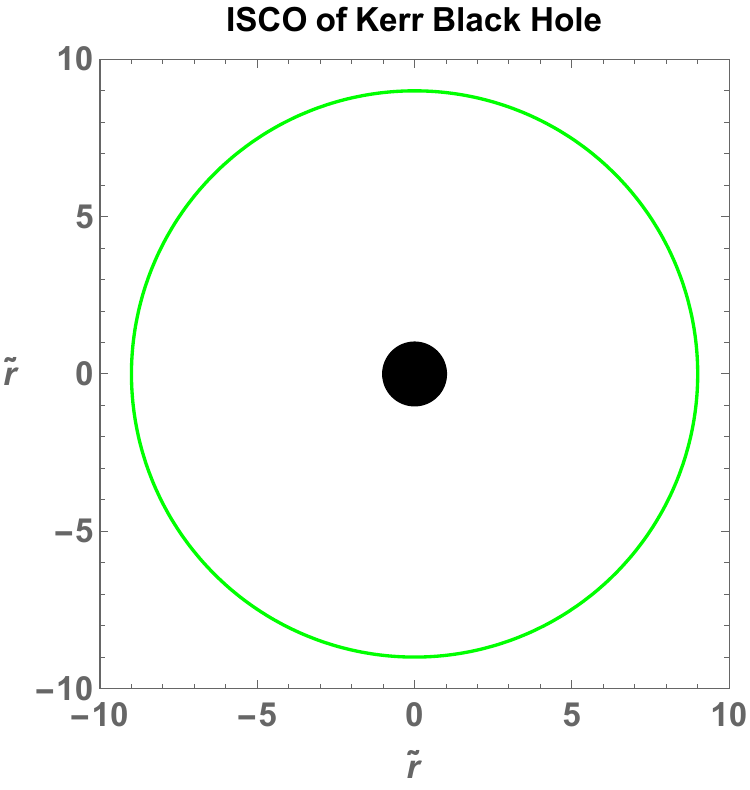}
{(c)}\includegraphics[width=5cm]{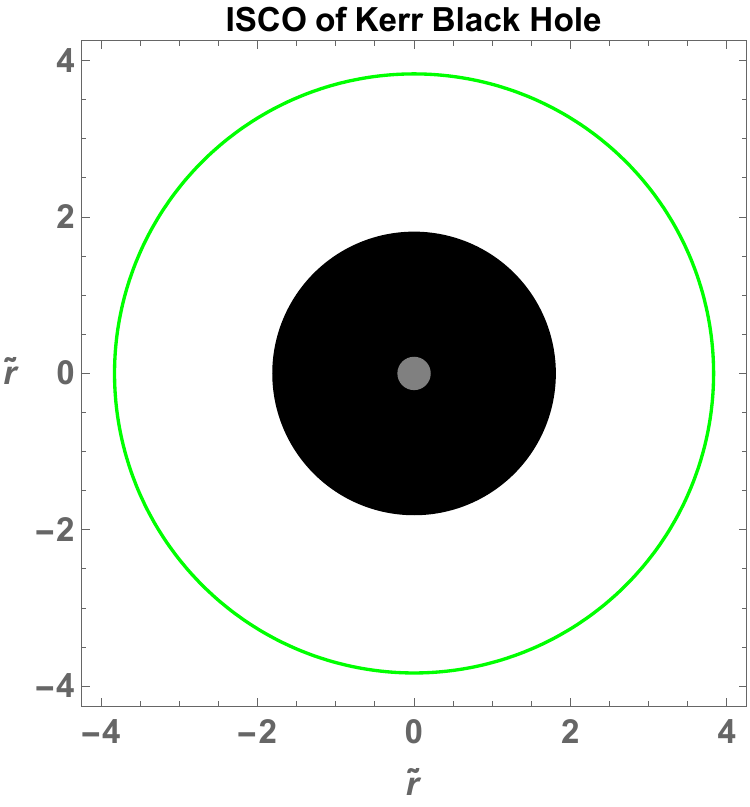}
\caption[figs]
{Innermost Stable Circular Orbits of Schwarzschild and Kerr black holes. (a) The black region is the event horizon of the Schwarzschild black hole ($\tilde{r}_{s}=2$ in dimensionless condition) and the green circle is the ISCO ($\tilde{r}_{ISCO}=6$) in dimensionless condition. (b) The black region is the event horizon ($\tilde{r}_{K}=1$ in dimensionless condition). The green circle is the ISCO ($\tilde{r}_{ISCO}=9$) in dimensionless condition. All these regions are shown by considering the $\tilde{a}=-1$ (retrograde orbit). (c) The gray region is the inner horizon ($\tilde{r}_{-}=0.2$ in dimensionless condition) and the black region is the event horizon ($\tilde{r}_{+}=1.8$ in dimensionless condition). The green circle is the ISCO ($\tilde{r}_{ISCO}=3.83$) in dimensionless condition. All these regions are shown by considering the $\tilde{a}=0.6$ (direct orbit).}
\label{fig41}
\end{figure*}

Furthermore, we show the ISCOs for Kerr and Schwarzschild black holes in Fig. \ref{fig41}, schematically with some details.

In Table \ref{tab:table1}, we present the locations of the event horizons and ISCOs in dimensionless form for various black hole models, including the Schwarzschild black hole, Kerr black hole, CDM black hole, and SFM black hole. The table covers two types of orbital configurations: retrograde motion and direct motion. This compilation allows for a straightforward comparison of the properties of these black holes and their ISCOs.

\subsection{CDM and SFDM Black Holes}

\begin{figure}
\centering
\includegraphics[width=6cm]{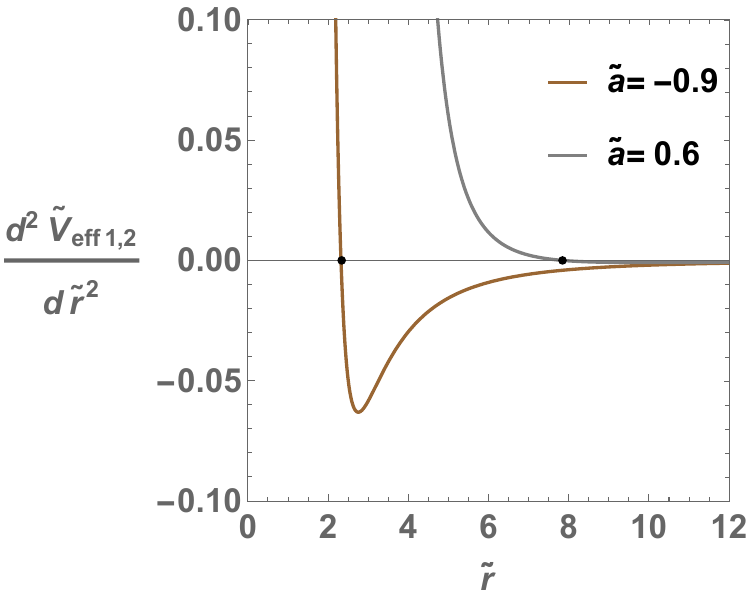}
\caption[figs]
{The second derivative of the effective potentials of the rotating black holes surrounded by cold dark matter halo and scalar field dark matter halo shows the locations of the ISCOs for two cases, $\tilde{a}=-0.9$ (Retrograde), and $\tilde{a}=0.6$ (Direct), by considering the values of $\tilde{R}_{s}=2.78 \times 10^{10}$, and $\tilde{\rho}_{c}= 0.50 \times 10^{-29}$ for CDM and $\tilde{R}=1.43 \times 10^{10}$, $\tilde{\rho}_{c}=2.81 \times 10^{-29}$ for SFDM.}
\label{fig430}
\end{figure}

\begin{figure}
\centering
{(a)}\includegraphics[width=5cm]{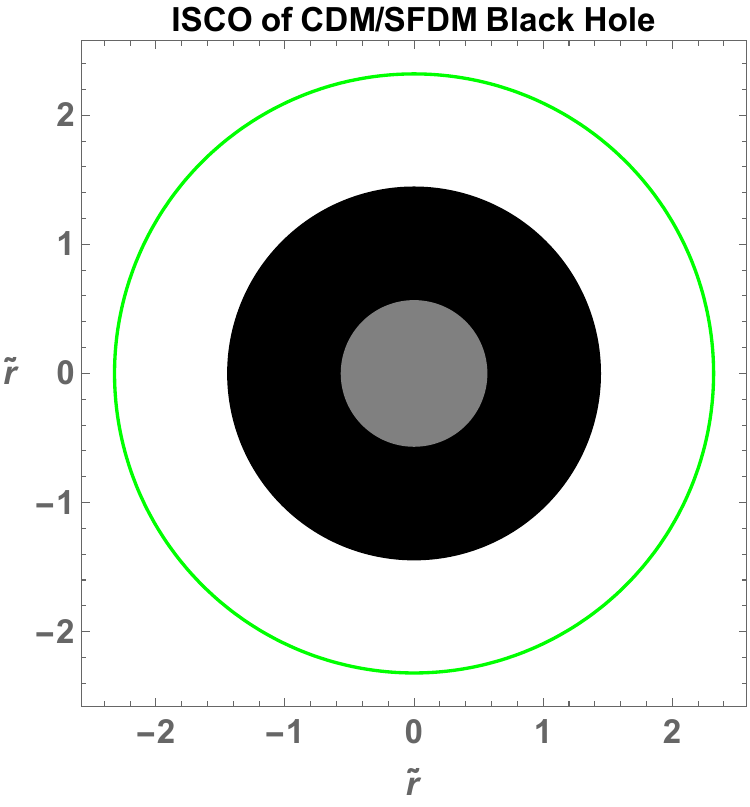}
{(b)}\includegraphics[width=5cm]{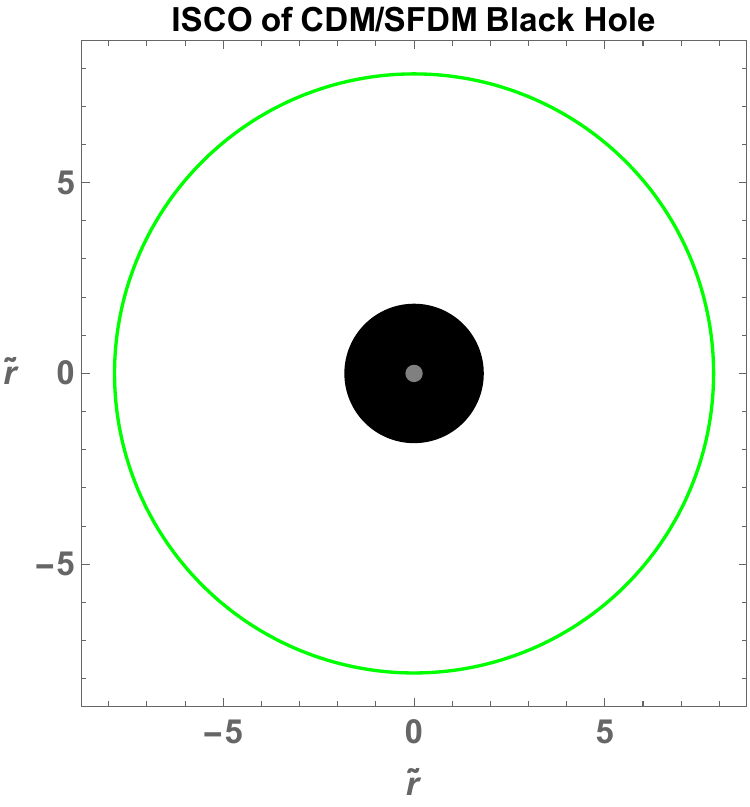}
\caption[figs]
{Innermost Stable Circular Orbits of rotating black holes surrounded by cold dark matter halo and scalar field dark matter halo. (a) The retrograde case, $\tilde{a}= -0.9$. (b) The direct case, $\tilde{a}=0.6$.}
\label{fig43}
\end{figure}

\begin{figure}
\centering
\includegraphics[width=9cm]{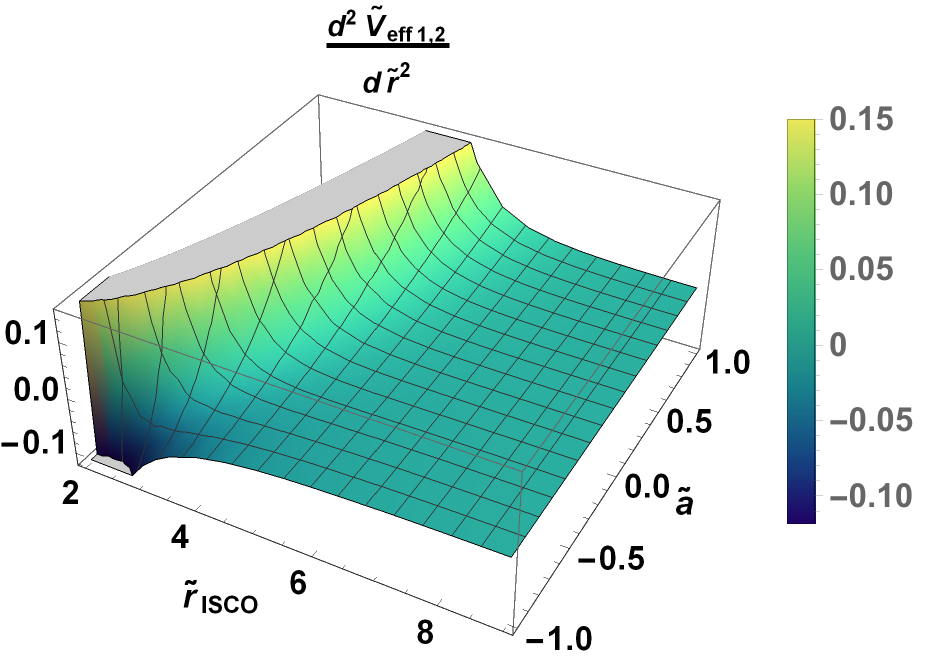}
\includegraphics[width=9cm]{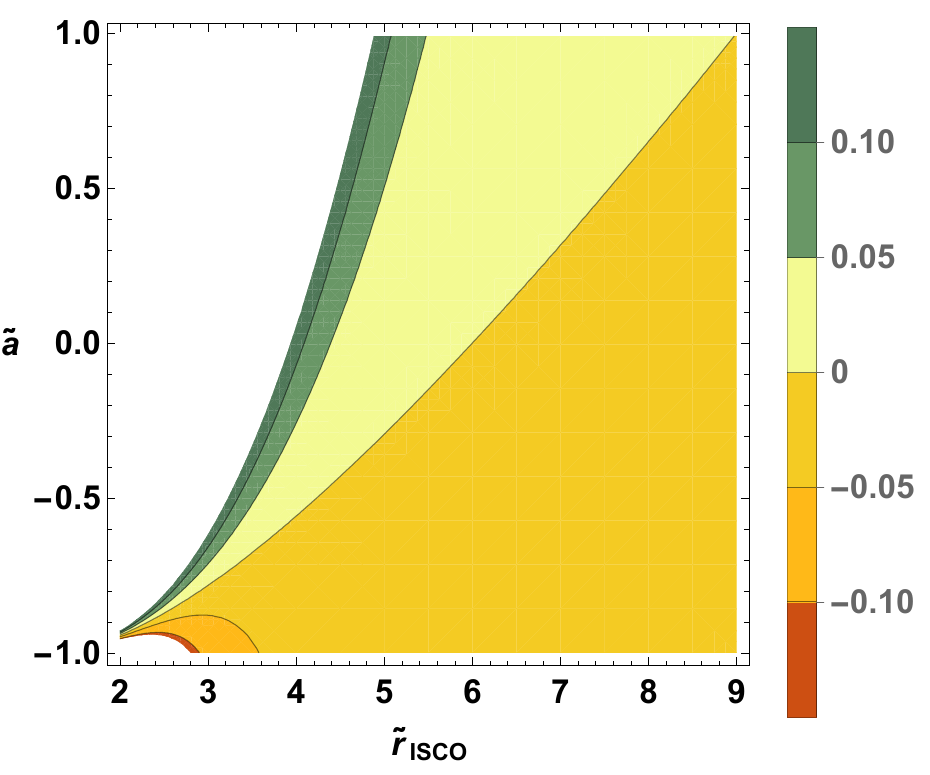}
\caption[figs]
{The range of $\tilde{r}_{ISCO}$ for varying values of $\tilde{a}$ (with 
$\tilde{a} < 0$ indicating retrograde motion, $\tilde{a}> 0$ indicating direct motion, and $\tilde{a} =0$ indicating non-rotating motion) can be determined by considering the points where the second derivative of the effective potential with respect to $\tilde{r}$ becomes zero (i.e., $\frac{d^{2}\tilde{V}_{eff 1, 2}}{d\tilde{r}^{2}}=0$) for rotating black holes surrounded by a cold dark matter halo and a scalar field dark matter halo. The figure encompasses both the CDM and SFDM cases within a single plot. The color schemes on the right-hand side of the figure correspond to different values of $\frac{d^{2}\tilde{V}_{eff}}{d\tilde{r}^{2}}$, with distinct colors indicating their respective ranges on the contour plots. These graphs are performed using the values of $\tilde{R}_{s}=2.78 \times 10^{10}$, $\tilde{\rho}_{c}=0.5 \times 10^{-29}$ for CDM, and $\tilde{R}=1.43 \times 10^{10}$, $\tilde{\rho}_{c}=2.81 \times 10^{-29}$ for SFDM.}
\label{fig8}
\end{figure}

In this step, demonstrated in Figure \ref{fig430}, we investigate the ISCOs for the rotating black holes surrounded by the cold dark matter halo and the scalar field dark matter halo, considering two cases: retrograde motion with $\tilde{a}=-0.9$ and direct motion with $\tilde{a}=0.6$. To obtain the locations of the ISCOs, we set the second derivative of the effective potentials to zero, utilizing Equations (\ref{V1}) and (\ref{V2}).
The resulting second derivative of the effective potentials with respect to $\tilde{r}$ is plotted in Figure \ref{fig430} for both cases of CDM and SFDM. Notice that their results are completely similar. 
The figure reveals that the ISCOs for the retrograde and direct cases are located at different positions compared to the Kerr black hole. Specifically, the ISCOs for retrograde motion are found at $\tilde{r}_{ISCO}=2.32$ for CDM and SFDM, while the ISCOs for direct motion are located at $\tilde{r}_{ISCO}=7.85$ for both cases of CDM and SFDM. Figure \ref{fig43} schematically shows the ISCOs of the rotating black holes surrounded by the cold dark matter halo and scalar field dark matter halo. These values indicate that the ISCOs decrease compared to the Kerr black hole. These results suggest that the presence of dark matter affects the stability of circular orbits near the black hole, leading to differences in the location of the ISCOs.
In Figure \ref{fig8}, we have also shown the ISCOs of the rotating black holes surrounded by the cold dark matter halo and the scalar field dark matter halo for different values of $\tilde{a}$.

As we pointed out in Table \ref{tab:table1}, the ISCOs of CDM and SFDM are the same by considering the selected values for $\tilde{R}_{s}$, $\tilde{R}$ and $\tilde{\rho}_{c}$.  We also have shown this issue in Figures \ref{fig430}, \ref{fig43}, and \ref{fig8}.

\subsection{Constraints from Sagittarius A*}

We now turn our attention to imposing constraints from the supermassive black hole at the Galactic Center of the Milky Way, Sagittarius A*. Notably, the diameter of the ISCO of Sagittarius A* is estimated to be around $10$ light minutes \cite{GRAVITY:2018ofz,Witzel:2018kzq,Doeleman:2008qh}, which implies a radius of approximately $5$ light minutes. Given the total mass of Sagittarius A* is $4.3$ million solar masses \cite{GRAVITY:2023avo}, we can calculate its ISCO radius using its Schwarzschild radius, which yields $7.14$ times its own Schwarzschild radius, or it could be written $7.14 (\frac{2G M_{S*}}{c^{2}})$(By using $GM_{S*}/c^{2}$, one transform of obtained dimensionless parameters into values with dimension, from Equation (\ref{Dm})). We then substitute this value into the second derivative effective potential equations for the two models and apply constraints on the dark matter halo parameters. Specifically, we set the ISCO radius equal to $\tilde{r}$ in the equation $\frac{d^{2}\tilde{V}_{eff}}{d\tilde{r}^{2}}$ and determine the allowed values of $\tilde{R}_{s}$, $\tilde{R}$ and $\tilde{\rho}_{c}$ for which the equation becomes zero, i.e., $\frac{d^{2}\tilde{V}_{eff}}{d\tilde{r}^{2}}=0$. Our goal is to identify the permissible parameter ranges for the dark matter halo in our models given the ISCO radius of Sagittarius A*. 

\subsubsection{CDM Black Hole}

In the first case, one can set the ISCO radius of Sagittarius A* equal to $\tilde{r}$ into the second derivative effective potential equation of CDM case $\frac{d^{2}\tilde{V}_{eff 1}}{d\tilde{r}^{2}}$ and determine the allowed values of $\tilde{R}_{s}$, and $\tilde{\rho}_{c}$ for which the equation becomes zero, i.e., $\frac{d^{2}\tilde{V}_{eff 1}}{d\tilde{r}^{2}}=0$.

\begin{figure}
\centering
\includegraphics[width=8cm]{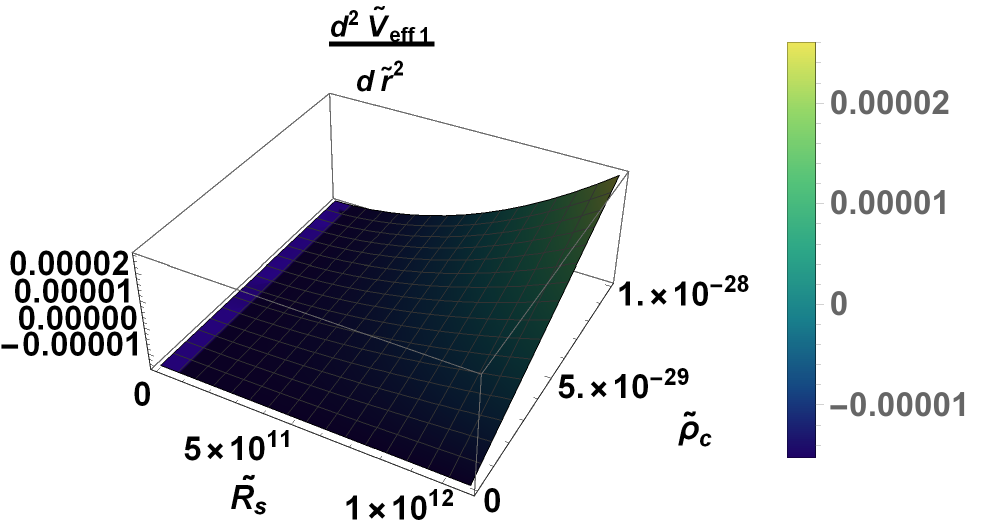}
\includegraphics[width=8cm]{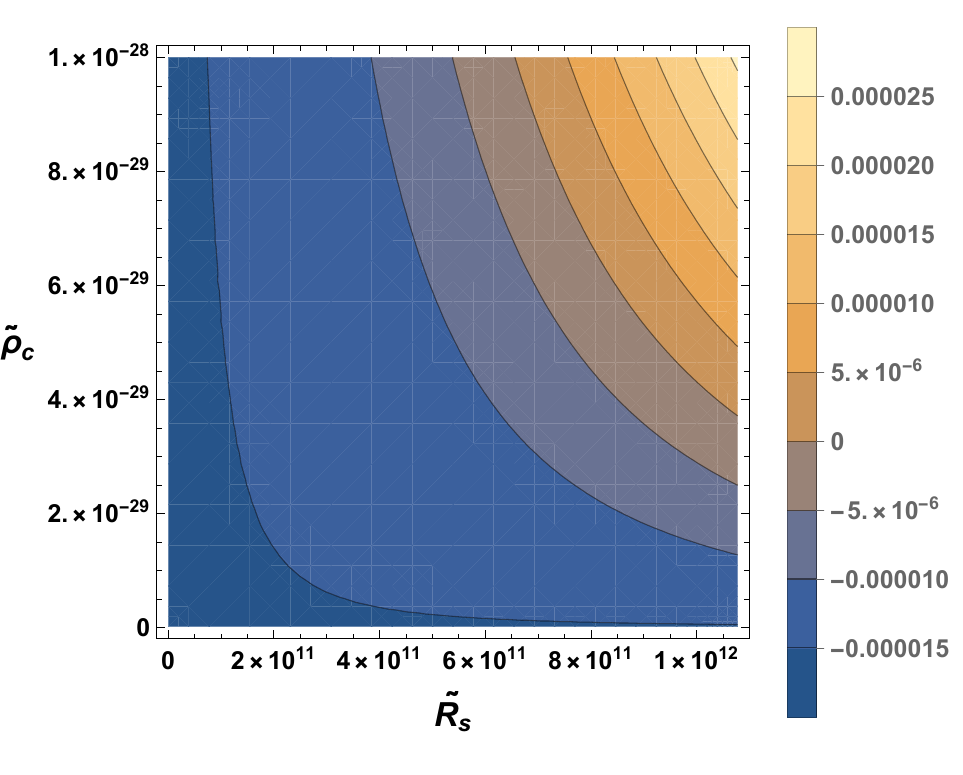}
\caption[figs]
{The acceptable ranges for $\tilde{R}_{s}$ and $\tilde{\rho}_{c}$ in the case of the rotating black hole surrounded by a cold dark matter halo yield the ISCO radius of Sagittarius A*. Specifically, for the direct case $\tilde{a}=0.36$. The color schemes presented on the right-hand side of the figure correspond to distinct values of $\frac{d^{2}\tilde{V}_{eff 1}}{d\tilde{r}^{2}}$, with each color representing a specific range on the contour plots.}
\label{fig12}
\end{figure}

In Figure \ref{fig12}, we have demonstrated the possible parameter ranges for the cold dark matter halo given the ISCO radius of Sagittarius A*. Moreover, we have presented these ranges for direct motion $\tilde{a}=0.36$.

The acceptable regions for $\tilde{R}_{s}$, and $\tilde{\rho}_{c}$ are depicted in the accompanying figures. Observing the graphs, one can note an inverted association between $\tilde{R}_{s}$, and $\tilde{\rho}_{c}$, whereby an increase in $\tilde{R}_{s}$ leads to a decrease in the permissible values of $\tilde{\rho}_{c}$ and vice versa.

To transform obtained dimensionless parameters into values with dimension, we should use $\frac{G M_{S*}}{c^{2}}$. As $r = \tilde{r} \times \frac{G M_{S*}}{c^{2}}$, it seems that to calculate $\rho_{c}$ in the dimension version, we should consider $\rho_{c} = \tilde{\rho}_{c} \times \frac{M_{S*}}{(\frac{G M_{S*}}{c^{2}})^{3}}$, so, we have $\rho_{c} = \tilde{\rho}_{c} \times (0.033 \times 10^{9}) \frac{kg}{m^{3}}$ or $\rho_{c} = \tilde{\rho}_{c} \times (4.86 \times 10^{26}) \frac{M_{\bigodot}}{p c^{3}}$. In addition, we can transform $R_{s}$ and $R$ by considering $R_{s} = \tilde{R}_{s} \times \frac{G M_{S*}}{c^{2}}$, and $R = \tilde{R} \times \frac{G M_{S*}}{c^{2}}$.

\subsubsection{SFDM Black Hole}

\begin{figure}
\centering
\includegraphics[width=8cm]{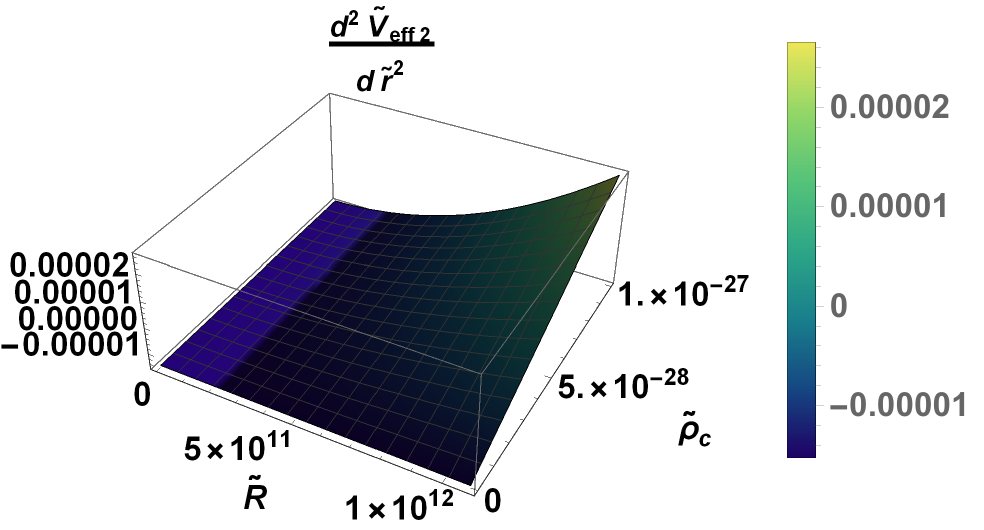}
\includegraphics[width=8cm]{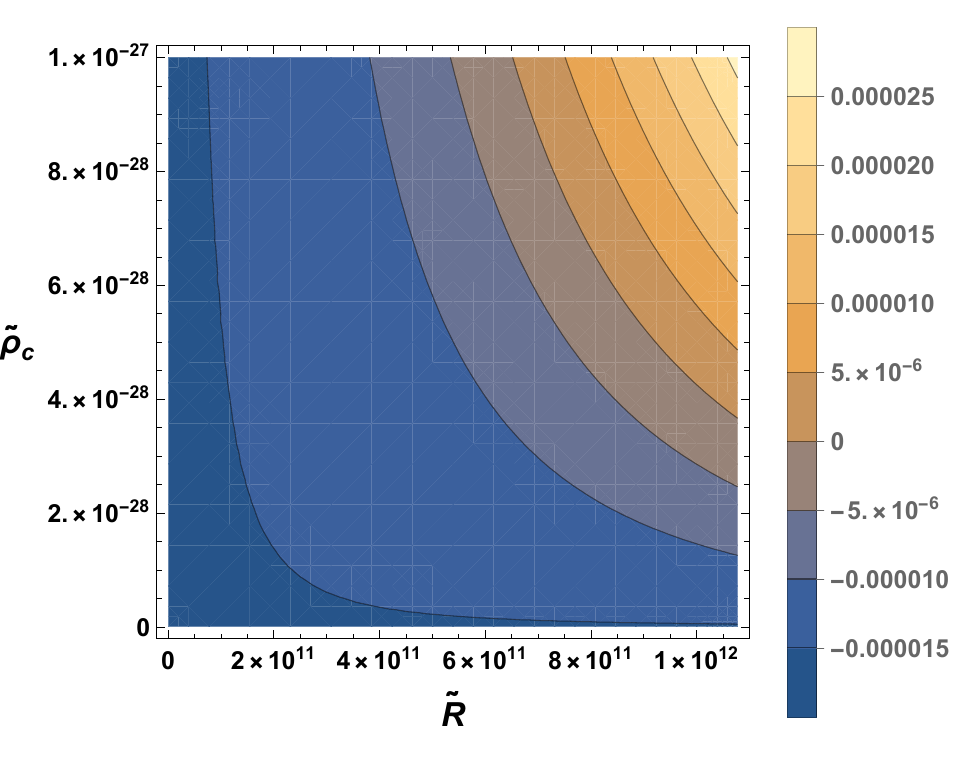}
\caption[figs]
{The acceptable ranges for $\tilde{R}$ and $\tilde{\rho}_{c}$ in the case of the rotating black hole surrounded by a scalar field dark matter halo yield the ISCO radius of Sagittarius A*. Specifically, for the direct case $\tilde{a}=0.36$. The color schemes presented on the right-hand side of the figure correspond to distinct values of $\frac{d^{2}\tilde{V}_{eff 2}}{d\tilde{r}^{2}}$, with each color representing a specific range on the contour plots.}
\label{fig13}
\end{figure}

To determine the allowed values of $\tilde{R}$, and $\tilde{\rho}_{c}$ for the scalar field dark matter halo, we can follow a similar approach by setting the ISCO radius of Sagittarius A* equal to $\tilde{r}$ in the second derivative effective potential equation of the SFDM case $\frac{d^{2}\tilde{V}_{eff 2}}{d\tilde{r}^{2}}$. By solving for zero, we find the permitted values of the halo parameters. This analysis is depicted in Figures \ref{fig13}, where we demonstrate the feasible parameter ranges for the SFDM halo given the ISCO radius of Sagittarius A* for $\tilde{a}=0.36$.

The acceptable ranges for $\tilde{R}$ and $\tilde{\rho}_{c}$ are depicted in the accompanying figures. Generally, observing the graphs reveals an interesting correlation between the two variables; as the value of $\tilde{R}$ increases, the permissible range of $\tilde{\rho}_{c}$ decreases, and vice versa.

As it is mentioned, to transform obtained dimensionless parameters into values with dimension, one can use $\frac{G M_{S*}}{c^{2}}$. In the CDM and SFDM cases, the acceptable ranges for $\tilde{R}_{s}$, $\tilde{R}$ and $\tilde{\rho}_{c}$ exhibit an inverted relationship. As $\tilde{R}_{s}$ and $\tilde{R}$ increases, the allowed values of $\tilde{\rho}_{c}$ decrease, and vice versa. 

The distinction between the CDM and SFDM cases lies in the nature of the dark matter halo and the resulting permissible parameter ranges. While the acceptable regions for $\tilde{R}_{s}$ and $\tilde{R}$ to achieve the ISCO radius of Sagittarius A* are the same in the two models, the acceptable values of $\tilde{\rho}_{c}$ exhibit differences. According to the obtained graphs, the acceptable values of $\tilde{\rho}_{c}$ in the CDM case lie within the range of $\tilde{\rho}_{c}=2 \times 10^{-29}$ to $\tilde{\rho}_{c}= 10^{-28}$. In contrast, the acceptable values of $\tilde{\rho}_{c}$ in the SFDM case fall within the range of $\tilde{\rho}_{c}= 2 \times 10^{-28}$ to $\tilde{\rho}_{c}= 10^{-27}$.
At the final of this stage, it is worth pointing out that the possible ranges for $\tilde{R_{s}}$, $\tilde{R}$ and $\tilde{\rho}_{c}$ are completely in agreement with other researchers \cite{Fernandez-Hernandez:2017pgq,Robles:2012uy}.

\section{Conclusions}\label{sec:6}

This work investigates the thin accretion disks around rotating black holes in spacetimes surrounded by cold dark matter halo and scalar field dark matter halo. Specifically, we study distinctions between these black holes and conventional cases, namely Schwarzschild and Kerr black holes.

Our analysis begins with an examination of the event horizons of the rotating black holes surrounded by dark matter halos, then the derivation of the equations of motion and effective potential for both geometries. We then computed the specific energy, specific angular momentum, and angular velocity of particles traversing circular orbits around the rotating black holes engulfed by cold dark matter halo and scalar field dark matter halo. Notably, we have demonstrated how the flux of radiant energy relates to the disk.

We plot the second derivative of the effective potential with respect to the radial coordinate, ensuring that the necessary condition for the existence of marginally stable orbits, $\frac{d^{2}\tilde{V}_{eff}}{d\tilde{r}^{2}}=0$, is met. We also present variations in specific energy, specific angular momentum, and angular velocity versus $\tilde{r}$.
Furthermore, we have pinpointed the locations of stable circular orbits and displayed them along with the event horizons in a table with different values of $\tilde{a}$. Notably, we recover the Schwarzschild black hole and its corresponding radius when $\tilde{a}=0$. Under certain circumstances, we have regained the event horizons and ISCOs of the Kerr black hole as well. We have compared the locations of event horizons and ISCOs of rotating black holes surrounded by dark matter halos with Schwarzschild and Kerr black holes.

Moreover, we have considered the constraints from the supermassive black hole at the Galactic Center, Sagittarius A*.

Its ISCO radius is approximately $5$ light minutes \cite{GRAVITY:2018ofz,Witzel:2018kzq,Doeleman:2008qh}, and we found its ratio to the observed Schwarzschild radius. Substituting this value into the second derivative effective potential equations, we can put constraints on the dark matter halo parameters. Our goal was to identify the permissible parameter ranges for the dark matter halo in our models given the ISCO radius of Sagittarius A*.

\section*{Acknowledgements}
This work is supported by the National Key $R\&D$ Program of China (Grant No. 2023YFE0117200), and the National Natural Science Foundation of China (No. 12105013).



\end{document}